\documentclass{elsarticle}
\usepackage{wrapfig,lipsum,booktabs}
\usepackage{amssymb}
\usepackage[]{amsmath}
\usepackage{multirow}
\usepackage{url}
\usepackage{booktabs}
\usepackage{graphicx}
\usepackage{balance}
\usepackage{subcaption}
\usepackage[linesnumbered, ruled,vlined] {algorithm2e}
\usepackage{color,soul}

\newcommand{\red}[1]{\textcolor{black}{#1}}

\newcommand{\alg}{{\bf{RelRank}}}
\newcommand{\Ni}{({\em i})~}
\newcommand{\Nii}{({\em ii})~}

\begin{document}
\begin{frontmatter}
\title{Study of Methods for Abstract Screening in a Systematic Review Platform}
\author[First]{Tanay Kumar Saha\corref{cor1}}
\ead{tksaha@iupui.edu}
\author[Second]{Mourad Ouzzani}
\ead{mouzzani@hbku.edu.qa}
\author[Second]{Hossam Hammady}
\ead{hhammady@hbku.edu.qa}
\author[Second]{Ahmed K. Elmagarmid}
\ead{aelmagarmid@hbku.edu.qa}
\author[Third]{Wajdi Dhifli}
\ead{wajdi.dhifli@univ-lille2.fr}
\author[First]{Mohammad Al Hasan}
\ead{alhasan@iupui.edu}
\cortext[cor1]{Corresponding Author and this work was done while the author was at QCRI. }

\address[First]{Indiana University Purdue University Indianapolis, Indianapolis, IN 46202, USA.}
\address[Second]{Qatar Computing Research Institute,
Hamad-Bin Khalifa University, Doha, Qatar.}
\address[Third]{University of Lille, Faculty of pharmaceutical and biological sciences, EA2694, BioMathematics Lab, 3 rue du docteur Laguesse, 59006 Lille, France.}

\begin{abstract}
A major task in systematic reviews is abstract screening, $i.e.$, excluding hundreds or thousands of irrelevant citations returned from one or several database searches based on titles and abstracts. 
\red{Most of the earlier efforts on studying systematic review methods for abstract screening evaluate the existing technologies in isolation and based on the findings found in the published literature. 
In general, there is no attempt to discuss and understand
how these technologies would be rolled out in an actual systematic review system.}
In this paper, we evaluate a collection of commonly used abstract screening methods over a wide spectrum of metrics on a large set of reviews collected from Rayyan, a comprehensive systematic review platform. 
We also provide equivalence grouping of the existing methods through a solid statistical test. 
Furthermore, we propose a new ensemble algorithm for producing a $5$-star rating for a citation based on its relevance in a systematic review.

In our comparison of the different methods, we observe that almost always there is a method that ranks first in the three prevalence groups. 
However, there is no single dominant method across all metrics. 
Various methods perform well on different prevalence groups and for different metrics. 
Thus, a holistic ``composite'' strategy is a better choice in a real-life abstract screening system.
We indeed observe that the proposed ensemble algorithm combines the best of the evaluated methods and outputs an improved prediction. It is also substantially more interpretable, thanks to its $5$-star based rating.

\end{abstract}


\end{frontmatter}

\newcommand{\ra}[1]{\renewcommand{\arraystretch}{1.2}}
\section{Introduction}

Randomized controlled trials (RCTs) constitute a key component of medical research and it is by far the best way of achieving results that increase our knowledge about treatment effectiveness~\cite{Chalmers:81}. Because of the large number of RCTs that are being conducted and reported by different medical research groups, it is difficult for an individual to glean a summary evidence from all the RCTs. The objective of systematic reviews is to overcome this difficulty by synthesizing the results from multiple RCTs.

Systematic reviews (SRs) involve multiple steps. Firstly, systematic reviewers formulates a research question and search in multiple biomedical databases. Secondly, they identify relevant RCTs based on abstracts and titles (abstract screening). Thirdly, based on full texts of a subset thereof, assess their methodological qualities, extract different data elements and synthesize them. Finally, report the conclusions on the review question~\cite{Uman:11}. However, the precision of the search step of a SR process is usually low and for most cases it returns a large number of citations from hundreds to thousands. SR authors have to manually screen each of these citations making such a task tedious and time-consuming. 
Therefore, a platform that can automate the abstract screening process is fundamental in expediting the systematic review process. To fulfill this need, we have built Rayyan~\footnote{\url{https://rayyan.qcri.org/}}---a web and mobile application for SR~\cite{ouzzani2016rayyan}. 
\red{As of November 2017, Rayyan users exceeded 8,000 from over 100 countries conducting hundreds of reviews totaling more than 11M citations.}

\begin{figure}[t]
\centering
\includegraphics[width=12cm]{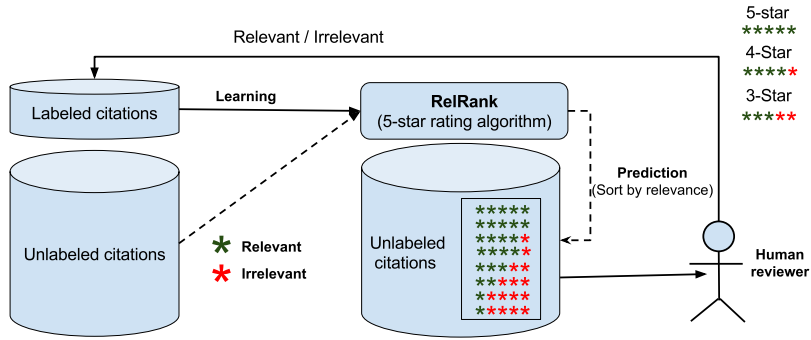}
\caption{Pictorial depiction of the abstract screening system Rayyan.}
\label{fig:pict-rayyan}
\vspace{-0.2in}
\end{figure}

\red{A primary objective of Rayyan is to expedite the abstract screening process. The first step of the abstract screening process is to upload a set of citations 
obtained from searches in one or more databases.
Once citations are processed and deduplicated, Rayyan provides an easy-to-use interface to the systematic reviewers so that they can label the citations as relevant or irrelevant. As soon as the systematic reviewer labels a certain amount of citations from both classes (relevant and irrelevant), Rayyan runs its 5-star rating algorithm and sorts the unlabeled citations based on their relevance. 
As the reviewer continues to label the citations, 
the system may trigger multiple runs of the 5-star rating algorithm. 
The process ends when all the citations are labeled. 
Thereof, one of the design requirements of the system is the ability to give feedbacks on the relevance of studies to users quickly. Figure \ref{fig:pict-rayyan} shows a pictorial depiction of our system.}

\red{Earlier efforts on studying systematic review methods in biomedical \cite{Alison.Thomas:15} and software engineering \cite{Olorisade.De:16} domains 
evaluate the existing technologies for abstract screening from the findings presented in the published literature. 
As pointed out by \cite{Alison.Thomas:15}, these evaluations do not make it clear how these technologies would be rolled out for use in an actual review system. In this study, we present 
our insights in designing methods 
(from both feature representation and algorithm design perspectives) for a light-weight abstract screening platform, Rayyan (Please see Section \ref{sec:method} for challenges, and Section \ref{sec:discussion} for study limitations).}

\red{A key difference of our evaluation, compared to existing studies such as \cite{Olorisade.De:16,Alison.Thomas:15}, is that our reviews come from an actual deployed systematic review system.  In addition, many of the results presented in the literature are hard to generalize as  they do not provide performance over multiple datasets and over different prevalence groups.
We thus collect a large set of reviews from 
the Rayyan platform and evaluate a large set of methods based on their practical utility to solve the class imbalance problem. 
Additionally, to give a clear insight to the class imbalance problem, we also divide the dataset into three prevalence groups, perform well defined statistical tests, and present our findings over a large set of metrics. 
The results of our evaluation give insights on the best approach for abstract screening in a real-life environment. 
Besides the evaluation of the existing methods, we also propose an ensemble method for the task of abstract screening that presents the prediction results through a 5-star rating scale. We make our detailed results and related resources available online\footnote{\url{https://tksaha.github.io/paper_resources/}}.}



\red{The primary aim of this study is to assess the performance of abstract screening task on the widely used SVM-based methods using a large set of reviews from a real abstract screening platform. This paper therefore addresses the following research questions from an abstract screening system design perspective:}

\begin{enumerate}
\item What are the challenges of using different feature space representations?
\item What kind of algorithms are suitable?
\item How SVM-based methods perform in different metrics over a large set of reviews and different class imbalance ratios?
\begin{itemize}
\item What aspects should be considered for designing a new algorithm for abstract screening?
\item Can an ensemble method improve performance?
\end{itemize}

\end{enumerate}

The remainder of this paper is organized as follows. In Section ~\ref{sec:rel_work}, we discuss related work. In Section~\ref{sec:method}, we provide an overview of the SVM-based methods used in the evaluation and details of the proposed $5$-star rating method. Section~\ref{sec:exp_result} presents the experimental results. In Section \ref{sec:discussion}, we discuss our key findings. We conclude in Section~\ref{sec:conclusion}.



\section{Related Work} \label{sec:rel_work}
We organize the related works into the following four groups: 
1)~works on ``abstract screening"  methods, 
2)~methodologies for handling ``data imbalance" in classification, as it is one of the main challenges in the abstract screening process,
3)~methodologies for ``active learning", a popular method for addressing data imbalance and user's labeling burden issue, and finally,
4)~works on ``linear review", from the legal domain, which bears some similarity with abstract screening in systematic reviews.
 
\subsection{Abstract Screening}
A large body of past research has  focused on automating the abstract screening process~\cite{Ambert:10,Cohen:08,Cohen.Ambert:12,Cohen.Hersh:06,Khabsa.Elmagarmid:15,Miwa.Thomas:14}. 
In terms of feature representation, most of the existing approaches~\cite{Cohen:08,Cohen.Hersh:06} use unigram, bigram, and MeSH (Medical Subject Headings).
An alternative to the MeSH terms is to extract LDA based latent topics from the titles and abstracts and use them as features~\cite{Miwa.Thomas:14}. Other methods~\cite{Cohen:08,Khabsa.Elmagarmid:15} utilize external information as features such as citations and co-citations. 
In terms of classification methods, SVM-based learning algorithms are commonly used~\cite{Ambert:10,Cohen:08,Cohen.Ambert:12}. 
According to a recent study~\cite{Olorisade.De:16}, $13$ different types of algorithms, including SVM, decision trees, naive Bayes, $k$-nearest neighbor, and neural networks have been proposed in existing works on abstract screening.  To the best of our knowledge, none of the existing works use the structured output version of SVM ($SVM^{perf}$~\cite{Joachims:05}) with different loss functions, which we have considered in this study. Note that $SVM^{perf}$'s learning is faster than a transductive SVM~\cite{Joachim:99}. For prediction, the former only keeps feature weights, which makes its prediction module very fast. On the other hand, transductive SVM takes both labeled and unlabeled citations into account, so its learning is slower than $SVM^{perf}$. 
\cite{Wallace.Kuiper:16} uses a  distant supervision technique to extract PICO (Population/Problem, Intervention, Comparator, and Outcome) sentences from clinical trial reports. 
The distant supervision method trains a model using previously conducted reviews and for the abstract screening task there is no such database for a particular review. 
So, we do not use distant supervision method. In this work, we evaluate various SVM-based classification methods with different loss functions over a set of abstract screening tasks.

\subsection{Data Imbalance}

Data imbalance in supervised classification is a well studied problem~\cite{Aggarwal:14,Joachim:99,Joachims:05,Shalev-Shwartz.Singer:11,Sun.Deng:12}. Two different kinds of approaches have been proposed to solve it. Methods belonging to the first kind focus on designing suitable loss functions: such as KL divergence~\cite{Esuli.Sebastiani:15}, quadratic mean~\cite{Liu.Chawla:11}, cost sensitive classification~\cite{Elkan:01}, mean average precision~\cite{Yue.Finley:07}, random forest~\cite{Chen.Liaw:04} with meta-cost, and AUC~\cite{Joachims:05}. Methods belonging to the second kind generate synthetic data for artificially balancing the ratio of labeled relevant and irrelevant citations. Examples of such methods are borderline-SMOTE~\cite{Han.Wang:05}, safe-level-SMOTE~\cite{Bunkhumpornpat.Sinapiromsaran:09}, and oversampling of the instances of the minority class along with under-sampling of the instances of the majority class. The authors in~\cite{Niculescu-Mizil.Caruana:05,Wallace.Dahabreh:12,Wallace.Dahabreh:14} use probability calibration techniques. 
In this paper, we evaluate the loss-centric approaches ignoring the data centric and probability calibration based approaches. The reason for our choice is the fact that synthetic data generation is computationally expensive which makes it very difficult to overcome extreme imbalance---a typical scenario for abstract screening. Also, for probability calibration we need a validation set which is not always available in a typical abstract screening session. For loss-centric approaches, we have modified $SVM^{perf}$ as proposed in~\cite{Esuli.Sebastiani:15,Liu.Chawla:11} to incorporate KL divergence and quadratic mean loss because the default implementation did not consider these loss functions.

\subsection{Active Learning}

As systematic reviews are done in batches, the problem of abstract screening can be modeled as an instance of batch mode active learning. Online learning trains the classifier after adding every citation whereas batch-mode active learning (BAL) does the same after adding batches of citations. However, BAL does not have any theoretical guarantee compared to online learning~\cite{Hanneke:14, Tong.Koller:00}. 
The task of BAL is to select batches with informative samples (citations) that would help to learn a better classification model. There are two popular methods to select samples: (1)~Certainty based and (2)~Uncertainty based. 
In certainty based methods, ``most certain" samples are selected to train the classifier while in uncertainty based method ,``most uncertain" ones are selected. In the uncertainty sampling-based methodologies, a large number of uncertainty metrics have been proposed; examples include entropy~\cite{Dagan.Engelson:95}, smallest-margin~\cite{Scheffer.Decomain:01}, least confidence~\cite{Culotta.McCallum:05}, committee disagreement~\cite{Seung.Opper:92}, and version space reduction~\cite{Nowak:09,Tong.Koller:00}. Using ``most uncertain" samples based on these metrics improves the quality of the classifier to find the best separating hyperplane, and thus improves the accuracy in classifying new instances~\cite{Miwa.Thomas:14}. On the other hand, certainty-based methods have also shown to be effective in carrying out active learning on imbalanced datasets, as demonstrated in~\cite{Fu.Lee:11}. In this paper, we present some interesting findings based on ``certainty" based sampling in abstract screening task.

\subsection{Linear Review}

A similar task to the abstract screening review is Technology Assisted Review (TAR)~\cite{Manfred.Paskach:13, Cormack:13,Grossman.Cormack:10, Henry:15,Roitblat.Kershaw:10,Saha.Hasan:15} which is popular among law firms. The main objectives of both review systems are the same, which is to classify between relevant results and irrelevant ones in a very imbalanced setup. TAR is used to prioritize attorneys' time to screen documents relevant to a lawsuit rather than those which are irrelevant. Generally, the number of documents are much larger (in millions) in TAR compared to the same in systematic review, so active learning is very popular in the TAR domain.
However, unlike systematic reviews, TAR researchers are interested in finding a good stabilization point where the training of the classifier should be stopped. Moreover, in TAR, achieving at least 75\% recall is considered as acceptable, whereas in systematic reviews 95\%-100\% recall is desired. Linear review faces similar set of challenges as is faced by abstract screening and in many times, they overcome those challenges by using similar solutions.

\section{Method} \label{sec:method}

For abstract screening, we have as input the title ($T_i$) and abstract ($A_i$) of a set of $n$ citations, $\mathcal{C} = \{C_{T_i,A_i}\}_{1 \le i \le n}$. We represent each citation, $C_{T_i,A_i}$ as a tuple $\langle \mathbf{x}_{C_i}, y_{C_i} \rangle$: $\mathbf{x}_{C_i}$ is a $d$-dimensional feature space representation of the citation and $y_{C_i} \in \{+1, -1\}$ is a binary label denoting whether $C_i$ is relevant or not for the given abstract screening task. For a labeled citation, $y_{C_i}$ is known and for an unlabeled citation it is not known. 
We use $\mathcal{L}$ and $\mathcal{U}$ to denote the set of  labeled and unlabeled  citations, respectively, 
$\mathcal{F}$ to represent features for a set of citations,
and $h$ for the hypothesis or hyperplane learned by training on $\mathcal{L}$. Some of our feature representation methods embed a word/term to a latent space; for a word $w$, we use $l(w) $ for the latent space representation of this word.

\subsection{\red{\bf Task Challenges}} \label{subsec:task-challenge}
The desiderata for supporting abstract screening in a SR platform like Rayyan includes the following: 
(1)~feature extraction should be fast and the features should be readily computable from the available data,
(2)~the learning and prediction algorithms should be efficient,
(3)~the prediction algorithm should be able to handle extreme data imbalance so that it can overcome the shortcomings of low search 
precision, and
(4)~the prediction algorithm should solve an instance of a two-class ranking problem such that the relevant citations should be 
ranked higher than the irrelevant ones. 
Based on these requirements, the choices 
for feature representations, the prediction
algorithms, and the prediction metrics are substantially reduced, as we discuss in the following paragraphs.

For feature extraction, we focus on two classes of methods namely uni-bigram and word2vec. Uni-bigram is a traditional feature extraction method for text data. Word2vec~\cite{Mikolov.Sutskever:13} is a recent distributed method which provides vector representation of words capturing semantic similarity between them. Both methods are fast and their feature representation is computed from the citation data which
are readily available. We avoided other feature extraction choices as they are either hard to compute or the information from which these
feature values are computed is not readily available. For example, features like co-citations are hard to obtain. 
Another possible feature is the frequency
of MeSH (Medical Subject Heading) terms. 
While MeSH  terms of a citation may be obtained from PubMed, this has some practical limitations especially in an automated system like Rayyan. In particular,  
(i)~PubMed does not necessarily cover all of the existing biomedical literature and
(ii)~to obtain MeSH terms, one has to either provide the PMID (PubMed Article Identifier) of each reference which is not always available, 
or use the title search API which is an error-prone process 
since it could return more than one result.
Thus, we avoided using this feature in our experiments. We also discarded the use of LDA (Latent Dirichlet Allocation) based feature generation and a matrix factorization based method \cite{kontonatsios2017semi} as an LDA based method is slow, and there is no easy way to set some of its critical parameters such as the number of topics. Methods based on distributed representations ($e.g.$, word2vec~\cite{Mikolov.Sutskever:13}) are shown to perform well in practice compared to the matrix factorization based alternatives. 
\red{Additionally, for distributed representation methods, Natural Language Processing (NLP) functions such as lemmatization (heuristic method to chop end of a word) and stemming (morphological analysis method to find root form of a word) are usually not a requirement as the methods automatically discovers the relationship between inflected word forms (for example, require, requires, required) based on the co-occurrence statistics and represents them closely in the vector space. For Uni-Bi (unigram, bigram) feature representation, one can use lemmatization or stemming. However, one immediate problem with lemmatization is that it is actually quite difficult to perform accurately for very large untagged corpora, so often a simple stemming algorithm (such as Porter Stemmer) is applied, instead, to remove all the most common morphological and inflectional word endings from the corpus \cite{Bullinaria2012}. However, we avoided such preprocessing step as it adds some extra cost to the preprocessing step and in our initial analysis, the performance difference was not statistically significant. }




\red{Since irrelevant citations are more common than the relevant ones, abstract screening algorithms suffer from the data-imbalance issue. Among the binary classification methods, the maximum-margin based methods ($aka$ SVM) have become very popular in the abstract screening domain~\cite{Alison.Thomas:15}. Thereof, we use SVM variants that are more likely to overcome the data-imbalance problem. Specifically, we use $SVM^{cost}$ (cost-sensitive SVM), and $SVM^{perf}$ (SVM with multivariate performance measure) with different loss functions for our evaluation.}

Finally, various prediction metrics have been used in the abstract screening domain which makes it difficult to compare among various models.
A recent survey of current approaches in abstract screening ~\cite{Alison.Thomas:15} reported that among the $69$ publications they surveyed, 
$22$ report performance using Recall, $18$ using  Precision, and $10$ using AUC. The non-uniform usage of these metrics makes it difficult to draw a conclusion about the performance of different methods. For example, if a particular work only reports Recall but not Precision, it is hard to understand its applicability.  
Among all the above metrics, the most common is
the Area Under the ROC Curve (AUC), which can be computed from the ranking of the citations as provided by the
classification model. Hence, to use AUC, the classification model for abstract screening needs to provide a ranked order of the citations. 
Among other ranking based metrics, AUPRC (area under Precision-Recall curve) is also used extensively. 
Moreover, authors of a recent work~\cite{Raeder:12} suggested to study the variability of the reported metrics and advocated to use a large number of repetitions ($500\times2$) to ensure reproducibility.
We also observed that most of the existing studies were performed on a small number of SRs, 
reported the evaluation with a different set of 
metrics, validated with a small number of repetitions, and most of them did not perform proper statistical significant tests
which would provide a statistical guaranty of the superiority of a method over the others.

In the following, we  describe the feature space representation, then the existing methods of automated abstract screening, and finally  our proposed $5$-star rating algorithm.

\subsection{Feature Space Representation}\label{sec:feat}
We use two types of feature  representation: (1) uni-bigram (Uni-Bi) and (2) word2vec (w2v). Uni-Bi based feature representation uses the frequency of uni-grams and bi-grams in a document. Note that uni-grams and bi-grams are generalization of $n$-grams, the collection of all contiguous sequences of $n$ words from a given text. Since the sum of the number of distinct words (uni-grams) and the number of distinct word-pairs (bi-grams)  over the document collection in a particular review task is generally large, Uni-Bi feature representation is high-dimensional. Besides, it produces a sparse feature representation, $i.e.$, a large number of entries in
the feature matrix is zero. Such high dimensional and sparse feature representation is poor for training a classification
model, so several alternative feature representations are proposed to overcome this issue. Among them, word2vec~\cite{Mikolov.Sutskever:13}
 is a popular alternative. It adopts a distributed framework to represent the words in a dense $d$-dimensional latent feature space. The feature vector of each word in this latent space is learned by shallow neural network models and the feature vector of
 a citation is then obtained by averaging the latent representation of all the words in that citation. 
It has been hypothesized that these condensed real-valued vector representation, learned from \emph{unlabeled} data, outperforms Uni-Bi based representations.

To learn vector representation of words using word2vec~\cite{Mikolov.Sutskever:13} for our corpus, we train the model on all abstracts and titles available in Rayyan. We use Gensim~\cite{Radim:10} with the following parameters: the number of context words as $5$, the min word count as $15$ and the number of dimensions in the latent space as $500$ ($d$=500, chosen through validation). Thus, for each word in the set of all available words ($w_i \in \mathcal{W}$), we learn a $500$-dimensional latent space representation. 
After averaging the latent vectors of words, we obtain the latent vector
of a citation on which we apply two kinds of normalization: (1)~instance normalization (row normalization) and (2)~feature based normalization (column normalization). These normalizations give statistically significant better results than using raw features for threshold agnostic metrics such as AUC and AUPRC. In instance normalization, we $z$-normalize the extracted features for each citation, $\mathbf{x}_{C_i}$.  For column normalization, we $z$-normalize in each of the $500$ dimensions. After both normalizations, we keep the feature values up to $2$-decimal places to minimize the memory requirement in our system.
Note that there exist some neural network based models such as sen2vec \cite{Le.Mikolov:14} which can learn the representation of a citation holistically instead of learning it by averaging the
representation of words in that citation. However, the sen2vec model is transductive in nature, $i.e.$, for new citations, we need to execute a few passes over the trained model, which is time-consuming. 

After learning the representation, for a limited number of words, we manually validate whether the latent space representation of a word captures the known semantic similarities of that words with various biomedical terms. Semantic similarities are computed using cosine similarity in the latent space following \cite{Mikolov.Sutskever:13}.
Our manual validation shows encouraging results. For example, the cosine similarity between ``liver" and ``cirrhosis" is $0.63$, which is large considering the fact that the vectors are of  500 dimensions. 
Also, for the query ``which words are related to {\em cirrhosis} in the same way {\em breast} and {\em cancer} are related", returns ``liver" as one of the top-$5$ answers in the w2v feature representation.

\begin{table}[t]
\centering
\begin{tabular}{@{}p{1.5cm}p{4cm}p{0.3cm}@{}}\toprule
{\bf{Feature}} & {\bf{Algorithm (Param.) }}  & {\bf {Id}}\\\midrule
\multirow{8}{*}{Uni-Bi} & $SVM^{perf}$ (b=0, AUC) & 1 \\
       & --- (b=1, AUC) & 2 \\
       & --- (b=1, KLD) & 3 \\
       & --- (b=1, QuadMean) & 4 \\
       & SVM (Default) & 5 \\
       & $SVM^{cost}$ (J, b=0) & 6\\
       & --- (J, b=1) & 7\\
       \cmidrule{2-3}
       & SVM Transduction & 11 \\ \hline 
\multirow{5}{*}{w2v row} & $SVM^{perf}$ (b=1, AUC)  & 21 \\
        & --- (b=1, KLD) & 22 \\
        & --- (b=1, QuadMean) & 23\\
        & $SVM^{cost}$ (J, b=0) & 24\\
        & --- (J, b=1) & 25\\ \hline 
\multirow{5}{*}{w2v col} & $SVM^{perf}$ (b=1, AUC)  & 31 \\
        & --- (b=1, KLD) & 32 \\
        & --- (b=1, QuadMean) & 33\\
        & $SVM^{cost}$ (J, b=0) & 34\\
        & --- (J, b=1) & 35\\
\bottomrule
\end{tabular}
\caption{Description of the algorithms used in the evaluation. We use the default regularization in all cases.  $J$ in $SVM^{cost}$ is set to the following ratio: $\frac{\text{\# of irrelevant citations in $\mathcal{L}$}}{\text{\# of relevant citations in $\mathcal{L}$}}$}
\label{tab:algo}
\end{table}

\subsection{Existing Algorithms}
A recent study \cite{Olorisade.De:16} reports that Support Vector Machine (SVM) is the most used algorithm in abstract screening. It is used in 31\% of the studies and at least one experiment annually since 2006. Moreover, as discussed in Section~\ref{sec:rel_work}, SVM-based methods are mostly used in both the data-imbalance and batch-mode active learning settings~\cite{Fu.Lee:11, Miwa.Thomas:14}. We thus restrict our evaluation to existing SVM-based algorithms.  SVM is a supervised classification algorithm which uses a set of labeled data instances and learns a maximum-margin separating hyperplane by solving a quadratic optimization problem. 
We should mention that the number of labeled citations can be very few at the start of a citation screening process and also the number of citations varies from a few hundred to a few thousand for different reviews. Thus, we did not try any supervised deep-learning based technique in our evaluation.

We use three types of SVM methods: (1)~inductive, (2)~transductive, and (3)~$SVM^{perf}$. Inductive SVM learns a hypothesis $h$ induced by $\mathcal{F}(\mathcal{L})$. Transductive SVM \cite{Joachim:99} reduces the learning problem of finding a $h \in \mathcal{H}$ to a different learning problem where the goal is to find one equivalence class from infinitely many induced by all the instances in $\mathcal{F}(\mathcal{U})$ and $\mathcal{F}(\mathcal{L})$. $SVM^{perf}$ exploits the alternative structural formulation of the SVM optimization problem for conventional binary classification with error rate~\cite{Joachims:06}. We use three different loss functions for the $SVM^{perf}$ implementation: (i)~AUC, (ii)~Kullback-Leibler Divergence (KLD), and (iii)~QuadMean Error~\cite{Liu.Chawla:11}. Table~\ref{tab:algo} shows the different SVM-based algorithms we used in our comparison along with their parameters and loss functions. We use the default regularization in all cases and $J$ in $SVM^{cost}$ is set to the following ratio: $\frac{\text{\# of irrelevant citations in $\mathcal{L}$}}{\text{\# of relevant citations in $\mathcal{L}$}}$. The ratio biases the hypothesis learner (the training algorithm) to penalize mistakes on the minority class (the relevant class) $J$ times more than the majority class (the negative class). In table \ref{tab:algo}, we assign distinct integer identifiers to represent each of the algorithms. For example, the first row in Table~\ref{tab:algo} refers to a method identified by Id $1$ that uses $SVM^{perf}$ with $b = 0$ and $\text{AUC}$ as the loss function. Comparison results among different methods (such as the results shown in Table~\ref{tab:rank-group}) are shown compactly by referring to each method by its identifier instead of its name. 

Making a comparison with the exact baseline algorithms proposed by various studies is not straightforward. 
31\% of the  studies reported in~\cite{Olorisade.De:16} use SVM as their classifier.
However, each one employs a different feature representation and a different SVM implementation. 
Thus, in this study we restrict our attention to a specific set of feature representations which are more suitable for our abstract screening platform, and use the linear kernel with different loss functions and the implementation provided by the author of each SVM algorithm.

\subsection{The Proposed $5$-star Rating Algorithm}\label{sec:5-star}
In our SR platform, we want to rank citations based on their graded relevance. 
The intuition is to help reviewers  better manage their time. To this end, we rate the citations from $1$ to $5$ using our proposed \alg\ algorithm. The citations with $5$ stars are relevant with high probability and need more attention whereas $1$ starred documents are irrelevant 
with high probability and may need less attention. Among the $5$ stars, we conceptualize $3-5$ stars to indicate relevant citations and $1-2$ stars for irrelevant ones.  

\begin{algorithm}[t]
\SetKwInOut{Input}{Input}\SetKwInOut{Output}{Output}
\SetKw{Ih}{ObtainInitialHyperplane}
\SetKw{SelectBatch}{SelectABatch}
\SetKw{QueryLabel}{QueryLabels}
\SetKw{Append}{Append}
\SetKw{List}{List}
\SetKw{Set}{Set}
\SetKw{Predict}{Predict}
\SetKw{GF}{GenerateFeature}
\SetKw{GC}{GenerateCombinedScore}
\newcommand\mycommfont[1]{\footnotesize\ttfamily\textcolor{black}{#1}}
\SetCommentSty{mycommfont}
\SetKwComment{Comment}{$\triangleright$\ }{}
\SetKw{Train}{Train}
\DontPrintSemicolon
\Input{$\mathcal{L}$, Labeled dataset; $\mathcal{U}$, Unlabeled dataset }
\Output{Score, $\mathcal{S}_{1 \le i \le |\mathcal{U}|}$}
\BlankLine
$\mathcal{F}_\mathcal{L}$, $\mathcal{F}_\mathcal{U} \leftarrow$ \GF ($\mathcal{L}$, $\mathcal{U}, \text{feature = W} $) \\
$h1 \leftarrow$ \Train ( $SVM^{perf}$, $\mathcal{F}_\mathcal{L}$)\\
$h2 \leftarrow$ \Train ( $SVM^{cost}$, $\mathcal{F}_\mathcal{L}$)\\
$\mathcal{S}_{h1} \leftarrow $ \Predict ($h1$, $\mathcal{F}_\mathcal{U}$) \\
$\mathcal{S}_{h2} \leftarrow $ \Predict ($h2$, $\mathcal{F}_\mathcal{U}$)\\
$\mathcal{F}_\mathcal{L}$, $\mathcal{F}_\mathcal{U} \leftarrow$ \GF ($\mathcal{L}$, $\mathcal{U}, \text{feature = U} $) \\ \tabularnewline 
$h3 \leftarrow$ \Train ( $SVM^{cost}$, $\mathcal{F}_\mathcal{L}$)\\
$\mathcal{S}_{h3} \leftarrow $ \Predict ($h3$, $\mathcal{F}_\mathcal{U}$) \\ \tabularnewline 
$\mathcal{S} \leftarrow$  \GC($\mathcal{U}$, $\mathcal{S}_{h1}$, $\mathcal{S}_{h2}$, $\mathcal{S}_{h3}$) \\
\Return {$\mathcal{S}$}
\caption{{\bf{RelRank}}: A Five Star rating algorithm using ensemble of max-margin based methods}
\label{alg:five-star-rater}
\end{algorithm}

Our $5$-star rating algorithm, \alg\ (Algorithm \ref{alg:five-star-rater}) uses an ensemble of SVM based methods. We choose method $SVM^{perf}$ (b=1, AUC) with w2v features, $SVM^{cost}$ (J, b=1) with w2v features, and $SVM^{cost}$ (J, b=1) with Uni-Bi features because of their special characteristics based on our initial evaluation of the methods on multiple datasets and over a range of metrics (described in Section \ref{sec:result:existing}). For instance, $SVM^{perf}$ (b=1, AUC) outperforms others based on the AUC and Recall metrics. When evaluated on the F1 metric (the harmonic mean of Precision and Recall), $SVM^{cost}$ (J, b=1) with w2v feature produces the highest value. On the other hand, $SVM^{cost}$ (J, b=1) with Uni-Bi feature has the highest Precision. So, in \alg, if all of these three methods agree on the relevance of a citation, then the citation gets $5$ stars; if two of them agrees, then it gets $4$ stars. Within a particular star, citations are ranked based on their average ranks (described in Algorithm~\ref{alg:gc}).

\begin{figure}[t]
\resizebox{0.95\textwidth}{!}{
\begin{algorithm}[H]
\SetKwInOut{Input}{Input}\SetKwInOut{Output}{Output}
\SetKw{Ih}{ObtainInitialHyperplane}
\SetKw{SelectBatch}{SelectABatch}
\SetKw{QueryLabel}{QueryLabels}
\SetKw{Append}{Append}
\SetKw{List}{List}
\SetKw{Set}{Set}
\SetKw{Predict}{Predict}
\SetKw{GF}{GenerateFeature}
\SetKw{GC}{GenerateCombinedScore}
\newcommand\mycommfont[1]{\footnotesize\ttfamily\textcolor{black}{#1}}
\SetCommentSty{mycommfont}
\SetKwComment{Comment}{$\triangleright$\ }{}
\SetKw{Train}{Train}
\SetKw{lambda}{lambda}
\SetKw{rank}{GetRanksForASetOfElements}
\DontPrintSemicolon
\Input{Unlabeled dataset($\mathcal{U}$), $\mathcal{S}_{h}[]$, $\mathcal{T}_{h}[]$, 
	Separation Threshold (ST), MAX Range (MR) }
\Output{Score, $\mathcal{S}$}
\BlankLine
$\text{U} =  |\mathcal{U}|$ \\ 
FVOTE = \lambda $\{|r| \text{exp}(-\frac{\text{U} - r}{\text{U}})\}$ \\
NORM = \lambda$\{|s| \frac{s - 1.0}{\text{U} -1.0} * \text{MR}\}$ \\
$r_{h}$ = \rank $(\mathcal{S}_{h})$ \\
\For {$u=0..|U|$}
{
	nv = $\mathcal{S}_{h[0][u]} \geqslant \mathcal{T}_{h[0]}\text{ ? 1 : -1 }$\\
	rs = $r_{h} [0][u]$\\
	fv = FVOTE (rs) \\
	\For {$v=1..|h|$}
	{
		$rs_{v}$ = $r_{h} [v][u]$	\\
		fv = fv + FVOTE ($rs_{v}$) \\
		rs = rs + $rs_{v}$	\\
		nv = $\mathcal{S}_{h[v][u]} \geqslant \mathcal{T}_{h[v]} ?
			(nv > 0 ? nv +1: nv) : (nv < 0 ? nv -1: nv)$ \\
	}
	ns = NORM ($rs/|h|$) \\
	finv = nv $\geqslant$  1 ? fv : nv \\
	$\mathcal{S}[u] $ = finv * ST + ns 
} 
\Return {$\mathcal{S}$}
\caption{GenerateCombinedScore}
\label{alg:gc}
\end{algorithm}
}
\end{figure}

In \alg, we first generate w2v row features for both sets $\mathcal{L}$ and $\mathcal{U}$. Then, we train $SVM^{perf}$ and $SVM^{cost}$ with w2v features 
to generate the first ($h1$) and second ($h2$) hyperplanes, respectively. We then compute scores for $\mathcal{U}$ based on the distances from $h1$ and $h2$. Using the Uni-Bi features with $SVM^{cost}$, we compute a third score $\mathcal{S}_{h3}$ using the hyperplane $h3$. Finally, we combine the three scores to generate a final score for $\mathcal{U}$ which we formally present in Algorithm~\ref{alg:gc} ({\bf {GenerateCombinedScore}}). 

\noindent {\bf {GenerateCombinedScore}} helps present the already classified citations in a ranked manner based on the scores 
obtained by calculating the distance from hyperplanes. The method takes the separation threshold (ST) and the max range (MR) as user-defined parameters. ST is used to define separation between ratings. MT is used for range normalization of a rank score between $1-\text{MR}$. Step~1 gets the number of unlabeled citations. Steps~2 and~3 define FVOTE and NORM as a lambda function. Both of them are used to transform ranks into a specified range. FVOTE smooths any value between $0.36-1.0$. In Step~4, we obtain ranks for each of the scores $\mathcal{S}_{1\le h \le |h|}$. 
The higher  the distance from the hyperplane for a particular citation is, the higher the rank. Steps $5$-$16$ calculate the final combined score. We iterate through all the scores in order. The very first hypothesis has the supreme power. If it predicts a particular citation as relevant, then the citation is rated between $3$ to $5$ star, and if it predicts it as irrelevant, then the same gets a $1$ or $2$ star. In Step 6, we get the vote from the dominant classifier, $i.e.$, if the score is greater than the classifier threshold then the vote count increases by $1$. The rank score and the fractional vote are obtained in Steps $7$ and $8$. The number of votes ($nv$) only increases if both the dominant classifier and the current hypothesis vote the citation as relevant. On the other hand, the number of votes decreases if both of them agree on its irrelevancy. Finally, in Step 14, we get a normalized rank between $1$ and $MR$. If the number of votes is greater than or equal to $1$ then the fractional vote is considered as the final vote otherwise the negative votes are used unchanged. This ensures that even if a particular citation gets positive votes, it still needs to get top rank to maintain its position in the rating. We use $1000$ and $800$ as values for the parameters ST and MR, respectively. We also use $0.0$, $2000$, and $2500$ as thresholds for \alg\ ($3$-star), \alg\ ($4$-star), and \alg\ ($5$-star), respectively, as these thresholds performed the best while performing
manual verification of system's performance.

\begin{table}[t]
\centering
\resizebox{0.95\textwidth}{!}{
\begin{tabular}{@{}llllcllllcllll@{}}\toprule
  \multicolumn{4}{c}{Prevalence [$0.22\%-5.92\%$]} &\phantom{abc}&
  \multicolumn{4}{c}{Prevalence [$6.79\%-13.07\%$]}&\phantom{abc}&
  \multicolumn{4}{c}{Prevalence [$13.45\%-40.08\%$]}\\
  \cmidrule {1-4} \cmidrule{6-9} \cmidrule{11-14}
  Review& Total & Rel. & Prev ($\%$) & \phantom{abc} & 
  Review& Total & Rel. & Prev ($\%$)& \phantom{abc} & 
  Review& Total & Rel. & Prev ($\%$)\\\midrule 
  P18 & 2241 & 5 & 0.22 & &
  P39 & 1149 & 78 & 6.79 & &
  P41 & 3250 & 437 & 13.45\\
  P31 & 3034 & 16 & 0.53 & &
  P16 & 484 & 34 & 7.02 & &
  P22 & 1352 & 193 & 14.28\\
  C12 & 1643 & 9 & 0.55 & &
  P32 & 895 & 63 & 7.04 & &
  P21 & 1352 & 193 & 14.28\\
  P5 & 8812 & 60 & 0.68 & &
  P19 & 541 & 39 & 7.21 & &
  P36 & 449 & 66 & 14.70  \\
  C9 & 1914 & 15 & 0.78 & &
  P44 & 643 & 50 & 7.78 & &
  P12 & 820 & 121 & 14.76 \\
  P30 & 1864 & 19 & 1.02 & &
  P23 & 565 & 49 & 8.67 & &
  P23 & 910 & 137 & 15.05\\
  P35 & 2601 & 33 & 1.27 & &
  P9 & 257 & 23 & 8.95 & &
  P20 & 2703 & 410 & 15.17\\
  C1 & 2544 & 41 & 1.61 & &
  C6 & 1113 & 100 & 8.98 & &
  P17 & 1704 & 266 & 15.61 \\
  P7 & 417 & 8 & 1.92 & &
  P27 & 1243 & 114 & 9.17 & &
  P38 & 1386 & 221 & 15.95 \\
  C5 & 1965 & 42 & 2.14 & &
  P6 & 2539 & 235 & 9.26 & &
  C4 & 899 & 146 & 16.24\\
  C2 & 845 & 20 & 2.37 & &
  P28 & 1242 & 115 & 9.26 & &
  P1 & 906 & 150 & 16.56 \\
  C13 & 3377 & 85 & 2.52 & &
  P43 & 996 & 95 & 9.54 & &
  P24 & 2488 & 419 & 16.84 \\
  P11 & 1580 & 41 & 2.59 & &
  P15 & 954 & 95 & 9.96 & &
  P42 & 4019 & 715 & 17.79 \\
  C14 & 660 & 24 & 3.64 & &
  P10 & 616 & 63 & 10.23 & &
  P2 & 1484 & 265 & 17.86\\
  P26 & 351 & 13 & 3.70 & &
  P33 & 640 & 68 & 10.63 & &
  P8 & 498 & 100 & 20.08 \\
  C11 & 1330 & 51 & 3.83 & &
  P46 & 551 & 59 & 10.71 & &
  C7 & 368 & 80 & 21.74 \\
  P4 & 1187 & 56 & 4.72 & &
  P34 & 1728 & 200 & 11.57 & &
  P45 & 957 & 230 & 24.03 \\
  C3 & 296 & 16 & 5.41 & &
  C8 & 343 & 41 & 11.95 & &
  C10 & 503 & 136 & 27.03 \\
  P37 & 730 & 40 & 5.48 & &
  P40 & 850 & 110 & 12.94 & &
  P3 & 1487 & 404 & 27.17 \\
  P25 & 338 & 20 & 5.92 & &
  C15 & 306 & 40 & 13.07 & &
  P14 & 822 & 256 & 31.14 \\
   &  &  &  & &
   &  &  &  & &
  P13 & 819 & 328 & 40.08 \\
\bottomrule
\end{tabular}}
\caption{Dataset Statistics. All publicly available reviews start with C (Cohen) whereas reviews from our system start with P (Private). Rel. stands for relevant citations whereas Prev. stands for prevalence of the dataset. For example, in dataset P18, the prevalence is 0.22\% as 5 (number of relevant citations) out of 2241 (total citations) is 0.0022.}
\label{tab:data-stat}
\end{table}

\section{Experimental Results} \label{sec:exp_result}
In this section, we first present our experimental settings, then we describe our dataset and performance metrics and finally present our experimental results.

\subsection{Experimental Settings}
We run all of our experiments on a computer with an Intel XEON 2.6Ghz processor running CentOS 6.7 operating system. For each dataset (described in the next section), we perform a $500\times2$ fold cross validation for each method to produce ``steady-state'' classifier performance following the recommendation from \cite{Raeder.Hoens:10}. In $n\times k$ fold cross validation, we split the data in $k$ blocks where the $k$-th block becomes the test block and the rest becomes the training data. We repeat this process $n$ times. The split is carried out through stratification. So for $500\times2$, we split each dataset into two blocks and then we use each block once as a training and once as a test. We repeat this process for $500$ times. According to \cite{Raeder.Hoens:10}, over the course of many iterations/repetitions (500 in our case) the average performance estimate for a given classifier may stabilize and produce ``steady-state'' classifier performance. It also produces substantially more repeatable results than using a small fixed number of iterations such as $5$ or $10$.

Now, we describe our algorithms parameter settings. For the inductive and transductive SVM methods, the default cost parameter, denoted as $c$, is computed as follows: 
the summation of all the $2$-norm values of the feature vectors are divided by the number of instances in $\mathcal{F}(\mathcal{L})$ to generate $b$. Finally, the fraction $\frac{1.0}{b*b}$ gives $c$. $J$ in $SVM^{cost}$ is set to the following ratio: $\frac{\text{\# of irrelevant}}{\text{\# of relevant}}$. Furthermore, we follow the recommendation for error loss function from~\cite{Joachims:05} to set the cost parameter for $SVM^{perf}$. The way we calculate the parameter values can also avoid the need for a representative validation set which is very hard to obtain in a systematic review platform. Note that for all the methods, we set the classification threshold to $0.0$ to simulate the actual behavior in our abstract screening platform.

\subsection{Datasets and Performance Metrics}\label{subsec:datasets_metrics}

We use $61$ reviews (dataset) for our experiments:
$15$ of them are publicly available from~\cite{Cohen:08}
and the other $46$ reviews were collected from Rayyan.
These represent 84K citations. In Table~\ref{tab:data-stat}, we present our dataset statistics. All publicly available reviews start with C (Cohen) whereas reviews from Rayyan starts with P (Private).  In the table, we give three statistics about each review: the total number of citations (Total), the total number of relevant citations (Rel.), and prevalence (ratio of relevant and total number of citations (Prev(\%))). The datasets of each group are sorted by their prevalence. For example, in the dataset P18, the prevalence is 0.22\% corresponding to the ratio of its $5$ relevant citations out of the $2241$ citations. We divide the reviews into three prevalence groups depending on the ratio of included citations: (1) Low ($0.22\%$ to $5.92\%$) (2) Mid ($6.79\%$ to $13.07\%$) and (3) High ($13.45\%$ to $40.08\%$). Generally, the smaller is the prevalence, the higher is the bias and the more difficult is the abstract screening task. The purpose of grouping the reviews into different prevalence groups is twofold: \Ni it allows to study the trend of various performance metrics within each group, $i.e$, depending on the complexity of the task (the smaller  the prevalence, the more complex the task), and \Nii it also makes our statistical tests robust against Type I error for data with frequent extreme values (the performance on some metrics varies wildly among different prevalence groups). 
\red{
We group reviews into three groups based on the severity of the class imbalance (high, mid, low) while  making sure that every group has a reasonable number of data samples (in our case, the number of datasets for each group is around 20).
Next, we describe our performance metrics.}


\begin{table}[ht]\centering
\ra{1.0}
\resizebox{1\textwidth}{!}{
\begin{tabular}{@{}p{2cm}p{8cm}p{6cm}@{}}\toprule
{\bf{Metric}} & {\bf{Definition}} & {\bf{Formula}} \\\midrule 
 Recall (Sensitivity) & Ratio of correctly predicted relevant citations to  all relevant ones.   & 
$\frac{\text{TP}}{\text{TP + FN}}$\\
Precision & Ratio of correctly identified relevant citations to all of those predicted as relevant.  
   & $\frac{\text{TP}}{\text{TP + FP}}$ \\
F-Measure  & Combines Precision and Recall values. It corresponds to the harmonic mean of Precision and Recall for $\beta = 1$. & $  \frac{ ({1 + {\beta}^2}).\text{Precision.Recall}} { {\beta}^2.\text{Precision + Recall}}$ \\
Accuracy  & Ratio of relevant and irrelevant citations predicted correctly to all citations.
 & $\frac{\text{TP + TN}} {\text{TP + TN + FP + FN}}$\\\tabularnewline 
\cmidrule{2-2}
ROC (AUC)  & Area under the curve obtained by graphing the true positive rate against the false positive rate; 1.0 is a perfect score and 0.5 is equivalent to a random ordering. &  \\
 AUPRC  & Area under precision recall curve.  &  \\\tabularnewline 
 \cmidrule{2-2}
AM ER. & Arithmetic mean of the loss in Recall of the relevant ($L_{R_p}$) and irrelevant class ($L_{R_n}$) & $\frac{\frac{\text{FN}}{\text{TP + FN}}  + \frac{\text{FP}}{\text{FP + TN}}  } {2}$\\ 
QD ER. & Quadratic mean, $aka.$ root mean square, measures the magnitude of varying quantities. It is defined as the square root of the arithmetic mean of the squares of the loss in Recall of the relevant ($L_{R_p}$) and the irrelevant class ($L_{R_n}$). & 
 $\displaystyle \sqrt {\frac{\frac{\text{FN}}{\text{TP + FN}} . \frac{\text{FN}}{\text{TP + FN}}  + \frac{\text{FP}}{\text{FP + TN}} . \frac{\text{FP}}{\text{FP + TN}}  } {2}}$
 \\\tabularnewline 
 \cmidrule{2-2}
Burden & The fraction of the total number of citations that a human must screen.  &  $\frac{\text{TP}^{L}+\text{TN}^{L}+\text{TP}^{U}+\text{FP}^{U} }{\text{N}}$\\
Yield  & The fraction of citations that are identified by a given screening approach. &  $\frac{\text{TP}^{L}+\text{TP}^{U}}{\text{TP}^{L}+\text{TP}^{U}+\text{FN}^{U}}$\\
Utility & It is a weighted sum of Yield and Burden. Here, $\beta$ is a constant. It represents the relative importance of Yield in comparison to Burden. We use $\beta=19$ to give 19 times more importance to yield in comparison to Burden in our experimental evaluations following the suggestion from~\cite{Wallace.Small:10}.
 & $\frac{\beta.\text{Yield} + (1 - \text{Burden})} {\beta + 1}$ \\
 \bottomrule
 \end{tabular}}
 \caption{Various metrics with their formulas and definitions. TP, FP, TN and FN represent true positive, false positive, true negative and false negative respectively.}
 \label{tab:metric}
 \end{table}

We use $11$ metrics for the evaluation~\cite{Alison.Thomas:15} (discussed in Table \ref{tab:metric}). The first four measures (Recall, Precision, F-measure and Accuracy) depend on a threshold and are widely used for evaluating automated abstract screening methods. In abstract screening, the highest cost is associated with false negatives (articles incorrectly identified as irrelevant) as these will not be manually reviewed and relevant evidence is omitted from the final decision (the systematic review). Therefore, it is expected that Recall should be very high in screening automation. AUC and AUPRC do not depend on thresholds and are common in binary classification problems with data imbalance. 
In an abstract screening platform, AUC and AUPRC are the second most important metrics. This is because a good ranking of documents makes it easier for the users to screen them, $i.e.$, they can mark a large batch of documents as relevant and irrelevant in a single operation. 
We also add two more metrics: 
(1)~Arithmetic mean error and (2)~Quadratic mean error. Both metrics measure the loss in Recall in both the relevant and irrelevant classes and thus are important for an  abstract screening platform. 
Finally, for active learning settings, we use Burden, Yield and Utility which respectively correspond to the fraction of the total number of citations that a human must screen, the fraction of citations that are identified by a given screening approach, and the weighted sum of Yield and Burden. 
From an abstract screening platform perspective, we want to minimize Burden and maximize Yield. 
 

\subsection{Statistical Test} \label{sec:exp_result:stat_test}
We have $18$ different methods (Table~\ref{tab:algo}) and 61 datasets. As we have partitioned the datasets into three prevalence groups: (1)~Low-prevalence, (2)~Mid-prevalence, and (3)~High-prevalence, we will compare the methods based on their overall performance on datasets of a particular prevalence group on a specific metric. Our goal is to generalize the findings on a larger population (dataset) that could fall within one of our defined prevalence groups. To perform a statistical test, we follow Cohen et $al$.~\cite{Cohen.McWeeney:07} who considered data and method as covariate to predict the performance metric, $i.e.$, \text{METRIC} $\sim$ \text{DATA} + \text{METHOD}.
To be more specific, the model resembles $y=mx+c$, where $x$ is a variable and $m,c$ are the constants. In this model, DATA and METHOD mimic $x$ and $c$ respectively. 
In our case, for a particular metric, we use the average of 500$\times$2 runs as the metric values for a particular method and a particular dataset, $i.e.$, multiple resampling from each dataset is to be used only to asses the performance score not its variance, which is similar in spirit to \cite{Janez:06}. The sources of variance are the differences in performance over (independent) datasets and not on (usually dependent) samples, so the elevated Type 1 error is not an issue.

We fit the model (\text{METRIC} $\sim$ \text{DATA} + \text{METHOD}) by using linear regression and perform a two-factor (DATA and METHOD) analysis of variance. This helps us in identifying whether there is any statistically significant difference among the methods, the datasets, and the interactions between the methods and the data. If the test shows a performance difference among the methods that is statistically significant, we again compare a pair of methods by using paired t-tests (post-hoc testing). 
Post-hoc testing with 18 methods leads to 18*17/2 possible post-hoc comparisons. Each of these 18*17/2 tests needs correction for multiple testings. To avoid the large number of pairwise testings, we follow these steps:

\begin{enumerate}
\item Find the best method and compare it with all the remaining methods via paired t-tests, and then utilize the Linear Step Up (LSU) procedure (also known as Benjamini and Hochberg procedure \cite{Benjamini.Hochberg:95}) to control the False Discovery Rate (FDR). It has been shown in \cite{Janez:06} that the Holm \cite{Hommel:88} and Hochberg~\cite{Benjamini.Hochberg:95} tests give practically equal results for post-hoc tests. 
\item \red{Separate the group of non-significant differences using LSU to perform FDR at the given alpha ($\alpha=0.05$). In our case, LSU \cite{benjamini2006adaptive} is performed using the following steps\footnote{\red{The exact p-values are used inside the LSU procedure to control the FDR. We provided only minimum and maximum p-values used for the rejection (fall into a different group) as the number of p-values generated from each pairwise comparison is enormous.}}:
\begin{enumerate}
\item Step 1: 
find the  first k  such that p(k) $\leq$ pos(k) $*$ $\alpha$/m. 
Here, m is the total number of methods that are tested against the best method in a rank group and p(1),...,p(k) are the corresponding ordered p-values, $i.e.$, p(1) $\geq$ p(2) $\geq$ p(3) ..... $\geq$ p(m). pos(k) is the position of a method in the ascending order of p-values. For example, pos(1) is m.
\item Step 2: if such a k exists, group first k-1 methods associated with p(1),...,p(k-1) in a single group with the best method as they do not have any statistically significant difference with the best method.
\end{enumerate}}
\item The group of methods from step 2 are isolated and the pairwise comparison is then repeated from step 1 with the remaining methods. The process continues until no more methods are left for pairwise comparison. 
\end{enumerate}

For example, if Step 2 yields 4 methods that are not worse than the best method by the desired statistically significant margin, then we will form our first group with the 5 methods (4+the best). The pairwise comparison will then be proceeded with the remaining 13 methods (18-5).

The above process yields equivalence groups of methods (which we call ranked groups, in short $rg$) such that within each group the performance of all the methods is statistically similar to the performance of the best method within that group. Also, across two groups the best method of a superior group is better than all the methods in the inferior group by a desired statistically significant margin.

\begin{table}[!t]
\centering
\resizebox{1\textwidth}{!}{
\begin{tabular}{@{}p{1.9cm}p{1cm}|p{1cm}|p{1cm}||
    p{1cm}|p{1cm}|p{1cm}||p{1cm}|p{1cm}|p{1.3cm}@{}}\toprule
  & \multicolumn{3}{c}{Prevalence} &
  \multicolumn{3}{c}{Prevalence}&
  \multicolumn{3}{c}{Prevalence}\\
  & \multicolumn{3}{c}{[$0.22\%-5.92\%$]} &
  \multicolumn{3}{c}{[$6.79\%-13.07\%$]}&
  \multicolumn{3}{c}{[$13.45\%-40.08\%$]}\\
  \cmidrule {2-4} \cmidrule{5-7} \cmidrule{8-10}
 {\bf{Metric}} & {$rg=1$} &{$rg=2$} & {$rg=3$}& 
  {${rg=1}$} &{$rg=2$} & {$rg=3$}&  
  {$rg=1$} &{$rg=2$} & {$rg=3$}\\\hline 
 {\bf{Precision}}  & \underline{\textbf{7}}, 11 & 3, \underline{\textbf{5}}, 25 & 4, 6, 35 &
    \underline{\textbf{5}},\underline{\textbf{7}} & 11 & 3, 6, 22, 25 & 
    \underline{\textbf{5}} &\underline{\textbf{7}} & 11 \\
       & ({\bf 40.05}) & ({\bf 25.82}) & ({\bf 22.78}) & ({\bf 57.86}) & ({\bf 46.40}) & ({\bf 39.36}) & ({\bf 73.83}) & ({\bf 59.91}) & ({\bf 55.26})  \\ \hline
 {\bf{Recall}} & \underline{\textbf{21}} & 31 & 1, 2, 34 &
    \underline{\textbf{21}} & 1, 2, 31, 34 & 22 & 
    \underline{\textbf{21}} & 1,2 & 31 \\
    & ({\bf 96.11}) & ({\bf 91.05}) & ({\bf 83.87}) & ({\bf 97.65}) & ({\bf 89.28}) & ({\bf 81.55}) & ({\bf 98.43}) & ({\bf 92.94}) & ({\bf 89.90}) \\ \hline
 {\bf{F-measure}}& \underline{\textbf{11}}, 25 & 24, 35 & 4, 22, 23 &
    \underline{\textbf{11}}, 25 & 4, 22, 23, 24, 35 & 3, 6, 33 & 
    7, \underline{\textbf{11}}, 25 & 35 & 24 \\
    & ({\bf 34.32}) & ({\bf 28.12}) & ({\bf 23.18}) & ({\bf 46.43}) & ({\bf 42.14}) & ({\bf 38.60}) & ({\bf 57.52}) & ({\bf 56.79}) & ({\bf 56.07}) \\ \hline
 {\bf{Accuracy}}&  \underline{\textbf{7}} & 5 & 3, 6, 11  &  
    5, \underline{\textbf{7}} &  11  & 6 & 
    5, \underline{\textbf{7}} &  11  & 3, 4, 6 \\
    & ({\bf 97.6}) & ({\bf 97.5}) & ({\bf 96.6}) & ({\bf 91.20}) & ({\bf 90.00}) & ({\bf 88.60}) & ({\bf 84.50}) & ({\bf 83.28}) & ({\bf 80.91}) \\ \hline 
 {\bf{AUC}}& \underline{\textbf{21}}, \underline{\textbf{25}} & 31, 35 & 1, 2, 5, 7, 23, 24 &
    \underline{\textbf{21}}, \underline{\textbf{25}} & 4, 24, 31, 35 & 1, 2, 5, 7, 22, 23 &
    \underline{\textbf{21}} & 7, 25 & (All Others) \\
 & ({\bf 86.30}) & ({\bf 86.26}) & ({\bf 85.13}) & ({\bf 85.99}) & ({\bf 83.54}) & ({\bf 83.10}) & ({\bf 86.24}) & ({\bf 86.00}) & ({\bf 84.91}) \\ \hline 
 {\bf{AUPRC}}& \underline{\textbf{1}}, \underline{\textbf{2}}, 5, 7, \underline{\textbf{21}} & 4, 25 & 11, 23, 24, 31 &
    \underline{\textbf{1}}, \underline{\textbf{2}}, 4, 5 & 7, 21 & 25 &
    \underline{\textbf{1}}, \underline{\textbf{2}} & (All Others) & - \\
  & ({\bf 33.15}) & ({\bf 32.53}) & ({\bf 32.15}) & ({\bf 47.64}) & ({\bf 46.78}) & ({\bf 45.80}) & ({\bf 61.71}) & ({\bf 57.75}) & - \\   
    \hline
 {\bf{AM ER.}} & \underline{\textbf{4}}, 31, 35, 24, \underline{\textbf{25}} & 1, 2, 11, 22, 23, 32, 33, 34 & 3, 6, 7, 21 &
    \underline{\textbf{4}}, \underline{\textbf{25}} & 24 & 31, 33, 35 & 
    \underline{\textbf{25}} & 24 & 4, 35 \\
  & ({\bf 0.29}) & ({\bf 0.32}) & ({\bf 0.41}) & ({\bf 0.243}) & ({\bf 0.248}) & ({\bf 0.270}) & ({\bf 0.217}) & ({\bf 0.227}) & ({\bf 0.238}) \\ \hline 
 {\bf{QD ER.}} & 1, 2, 4, \underline{\textbf{24}}, 25, 31, 32, 33 & 23, 34, 35  &   11, 22 &
    4, 25, 33 & \underline{\textbf{24}} & 23, 31, 35 & 
    25 & \underline{\textbf{24}} & 4, 35 \\
  & ({\bf 0.38}) & ({\bf 0.40}) & ({\bf 0.46}) & ({\bf 0.26}) & ({\bf 0.27}) & ({\bf 0.32}) & ({\bf 0.22}) & ({\bf 0.23}) & ({\bf 0.24}) \\ \hline  
 {\bf{Yield}}&\underline{\textbf{21}} & 31 & 1, 2, 34 & 
    \underline{\textbf{21}} & 1, 2, 31, 34  & 32 & 
    \underline{\textbf{21}} & 1, 2  & 31 \\
 & ({\bf 98.04}) & ({\bf 95.51}) & ({\bf 91.95}) & ({\bf 98.82}) & ({\bf 94.64}) & ({\bf 90.77}) & ({\bf 99.21}) & ({\bf 96.47}) & ({\bf 94.95}) \\ \hline    
 {\bf{Burden}} &\underline{\textbf{5}} & 7 & 3, 6 & 
    \underline{\textbf{5}} & 7  & 6 & 
    \underline{\textbf{5}} & 11  & 6 \\
 & ({\bf 50.06}) & ({\bf 50.26}) & ({\bf 50.66}) & ({\bf 50.70}) & ({\bf 52.45}) & ({\bf 54.01}) & ({\bf 53.72}) & ({\bf 59.06}) & ({\bf 61.43}) \\ \hline 
 {\bf{Utility}} & \underline{\textbf{21}}  & 31 & 1, 2, 34 & 
    \underline{\textbf{21}} & 1, 2, 31, 34  & 32 & 
    \underline{\textbf{21}} & 1,2  & 31 \\
  & ({\bf 93.64}) & ({\bf 91.97}) & ({\bf 88.62}) & ({\bf 94.32}) & ({\bf 91.03}) & ({\bf 87.59}) & ({\bf 94.60}) & ({\bf 92.56}) & ({\bf 91.44}) \\ \hline    
\bottomrule
\end{tabular}
}
\caption{\red{Grouped results of all the methods using metrics from Table~\ref{tab:metric} ($rg$ stands for rank group) along with representative values (within brackets). For easy of reading, we recall the methods as described in Table~\ref{tab:algo}.
Method 1 -- uni-bi feature with $SVM^{perf}$ with b=1, and AUC. 
Methods 2, 3, and 4 -- uni-bi feature with $SVM^{perf}$ with b=1, and AUC, KLD, and QuadMean loss functions, respectively. 
Method 5 --  SVM (Default).
Method 11 -- to SVM transduction. 
Methods 6 and 7 --  uni-bi feature with $SVM^{cost}$ with  b=0 and b=1, respectively.  
Methods 21, 22, and 23 --  w2v row feature with $SVM^{perf}$ with b=1, and AUC, KLD, and QuadMean loss functions, respectively. 
Methods 31, 32, and 33 --  w2v col feature with $SVM^{perf}$ with b=1, and AUC, KLD, and QuadMean loss functions, respectively.  
Methods 24 and 25 --  w2v row feature with $SVM^{cost}$ with b=0 and b=1, respectively. 
Methods 34 and 35 -- w2v col feature with $SVM^{cost}$ with  b=0 and b=1, respectively.
}} 
\label{tab:rank-group}
\end{table}


\begin{table}[!t]
\centering
\resizebox{1\textwidth}{!}{
\begin{tabular}{@{}l||c||c||c@{}}\toprule
&{Prevalence} &
  {Prevalence}&
  {Prevalence}\\
&{[$0.22\%-5.92\%$]} &
    {[$6.79\%-13.07\%$]}&
   {[$13.45\%-40.08\%$]}\\\hline 
& {\bf P-value} & {\bf P-value} & {\bf P-value} \\
Metric & (min., max.) & 
(min., max.) &
(min., max.) \\\midrule
Precision & (3.68e-07, 0.03) & (1.17e-14, 0.03) & (3.63e-14, 0.03)\\
Recall & (7.53e-22, 0.04)  & (1.24e-20, 0.007) & (2.09e-14, 0.02) \\
F-measure & (6.21e-09, 0.015) & (4.94e-12, 0.03)  & (2.12e-11, 0.03) \\
Accuracy & (6.33e-22, 0.02)  & (1.36e-21, 0.04)  &(1.98e-14, 0.04)  \\\hline 
AUC & (5.19e-09, 0.03) & (4.93e-11, 0.03)  & (1.58e-11, 0.03) \\
AUPRC & (1.44e-09, 0.03)  & (1.23e-10, 0.03)  & (1.44e-09, 0.03) \\\hline 
AM ER. & (1.26e-15, 0.03) & (1.76e-15, 0.04)  &  (9.75e-12, 0.016) \\
QD ER. & (1.15e-20, 0.03)  & (2.03e-16, 0.02) & (2.06e-12, 0.02) \\
\hline
Yield & (8.14e-22, 0.04)& (1.24e-20, 0.007)  & (2.09e-14, 0.02) \\
Burden & (2.39e-26 0.01) & (1.66e-25, 0.03) & (7.40e-16, 0.03) \\
Utility & (8.40e-22, 0.04)  & (1.66e-20, 0.006) & (3.23e-14, 0.02) \\
\bottomrule 
\end{tabular}}
\caption{\red{P-values of all the metrics corresponding to the Table \ref{tab:rank-group}. The statistics give some insights on how conservative we are in selecting two statistically insignificant methods to put into the same group (Please see Section \ref{sec:exp_result:stat_test} for details). For example, for the Precision metric, any p-value greater than 0.3 may put two methods into the same group for some rank group, even though we choose $\alpha=0.05$ in our case. This is because of the Linear Step Up (LSU) procedure described in \ref{sec:exp_result:stat_test}.} }
\label{tab:p-value-result}
\end{table}

\subsection{Results of Existing Baselines} \label{sec:result:existing}

Table~\ref{tab:rank-group} shows the comparison results. Along the rows and the columns, we list respectively the 11 performance metrics and the three prevalence groups (low, mid, and high). The methods that perform the best for a given metric within a prevalence group are organized within sub-columns of ranked group. The ranked groups of methods are created by a statistical test with $\alpha$=$0.05$. We only show the top-3 ranked groups which are organized under sub-columns titled $rg=1$, $rg=2$, and $rg=3$.  Within a ranked group, the methods are listed by their identifiers (as shown in Table~\ref{tab:algo}). \red{For each rank group, metric, and prevalence range, we report a representative metric value in brackets. Our detailed results are available online\footnote{\url{https://tksaha.github.io/paper_resources/}}.  We also provide the p-values of all the metrics corresponding to Table \ref{tab:rank-group} for our statistical test in Table \ref{tab:p-value-result}. The statistics give some insights on how conservative the LSU method is in selecting two methods to put into the same group (Please see Section \ref{sec:exp_result:stat_test} for details). For example, for the Precision metric, any p-value greater than 0.3 may put two methods into the same group for some rank group. This is because of the LSU procedure described earlier in Section \ref{sec:exp_result:stat_test}.} As a convenience to the reader, note that methods that use the Uni-Bi, w2v row, and w2v col features have identifiers in the range of $1-11$, $21-25$ and $31-35$ respectively.


\red{We first discuss the best methods by considering the metrics that depend on a threshold. Results for these metrics are shown in the first four rows and the last 5 rows of Table~\ref{tab:rank-group}. We will discuss the best methods by their names with their identifiers within parenthesis. 
For Precision, $SVM^{cost}$ ($7$) outperforms the other methods in the low and mid prevalence groups achieving around 40\% and 58\% for low and mid prevalence groups, respectively. But, for the high prevalence group, it falls in the second tier and obtains a Precision value of 60\%. SVM Default ($5$) also performs very well with a second tier rank in the low prevalence group and a first tier rank in the mid and high prevalence groups. SVM Default ($5$) achieves  around 25\%, 58\%, and 74\% in Precision for low, mid, and high prevalence groups, respectively. For Recall, w2v row with $SVM^{perf}$ (AUC) ($21$) is in the top position in all three prevalence groups obtaining around 96\%, 97\% and 98\% in Recall for low, mid, and high prevalence groups, respectively. $SVM^{trans}$ with Uni-Bi ($11$) and $SVM^{cost}$ with w2v row ($25$) are the top performers for the F1 metric. These methods achieve around 34\%, 46\%, and 58\% in F1 value for low, mid, and high prevalence groups, respectively. For Accuracy, $SVM^{cost}$ (b=1) with Uni-Bi ($7$) performs the best followed by $SVM^{trans}$ with Uni-Bi ($11$). $SVM^{cost}$ (b=1) with Uni-Bi ($7$) achieves 97\%, 91\%, and 85\% in Accuracy in low, mid, and high prevalence groups, respectively. The decreasing trend in performance values is because the lower is the prevalence the more the classifier is able to achieve higher accuracies by predicting the label of all the instances as the dominating class. $SVM^{cost}$ with w2v row normalized features, $24$ and $25$, rank first in Arithmetic Mean (AM) and Quadratic Mean (QM) Errors, respectively. Again, w2v row with $SVM^{perf}$ (AUC) ($21$) ranks first in the Yield and Utility metrics while the SVM Default ($5$) is at the top of the list in the Burden metric. $SVM^{perf}$ (AUC) ($21$) achieves around 98\% in Yield and 94\% in Utility for all the prevalence groups.}


\red{We now analyze the performance of the methods on other metrics which are threshold agnostic. For instance, in AUC, w2v row with $SVM^{perf}$ (AUC) ($21$) performs the best. $SVM^{cost}$ (b=1) with w2v row normalized features ($25$) also performs well. Both methods achieve around 86\% AUC in all three prevalence groups. In AUPRC, $SVM^{perf}$ with (b=0, AUC) ($1$) and (b=1, AUC) ($2$) along with the Uni-Bi features are the top performing methods. They obtain 33\%, 47\%, and 62\% in low, mid, and high prevalence groups, respectively. }

\begin{figure}[t]\centering
\resizebox{0.95\textwidth}{!}{
\begin{tabular}{@{}ccc@{}}\toprule
 {{\bf {Prevalence}}} &
  { {\bf {Prevalence}}}&
  { {\bf {Prevalence}}}\\
  { [$0.22\%-5.92\%$]} &
  {  [$6.79\%-13.07\%$]}&
  { [$13.45\%-40.08\%$]}								\\\midrule   
   \includegraphics[width=7cm]{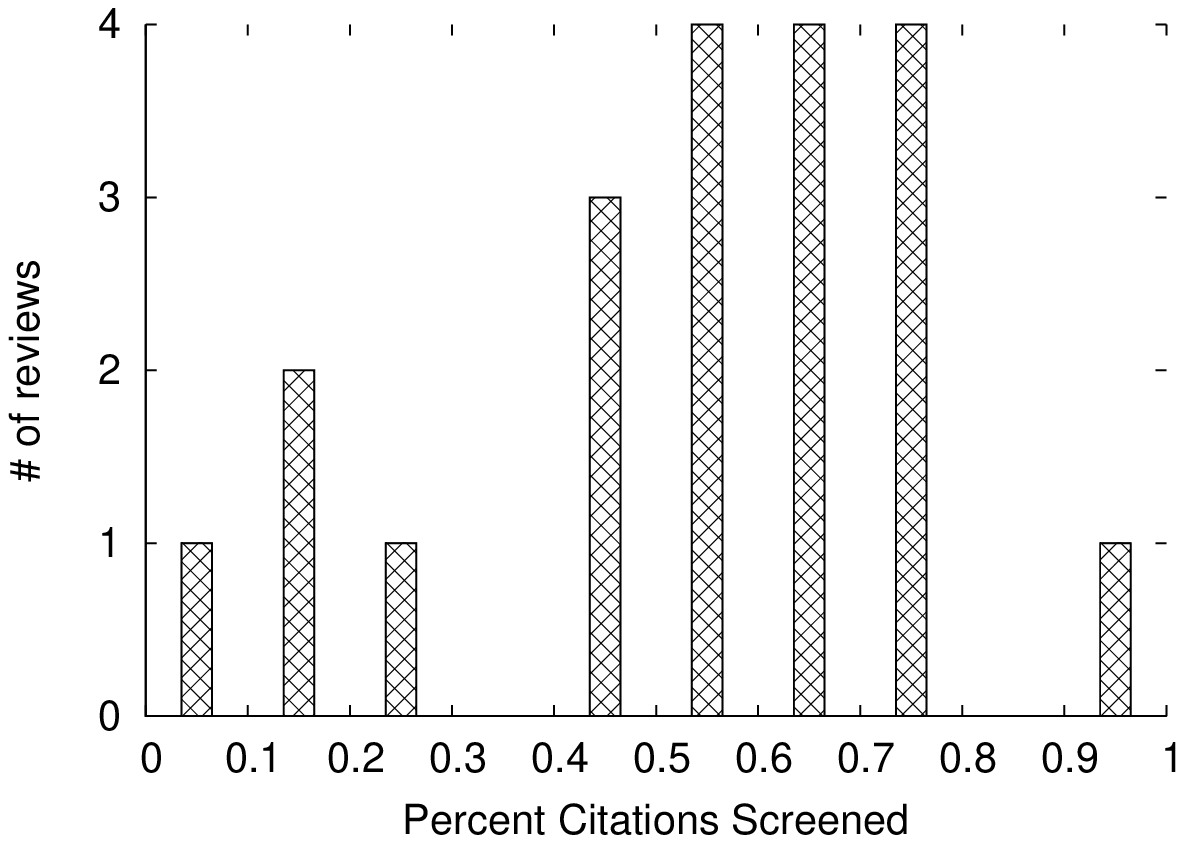} &
   \includegraphics[width=7cm]{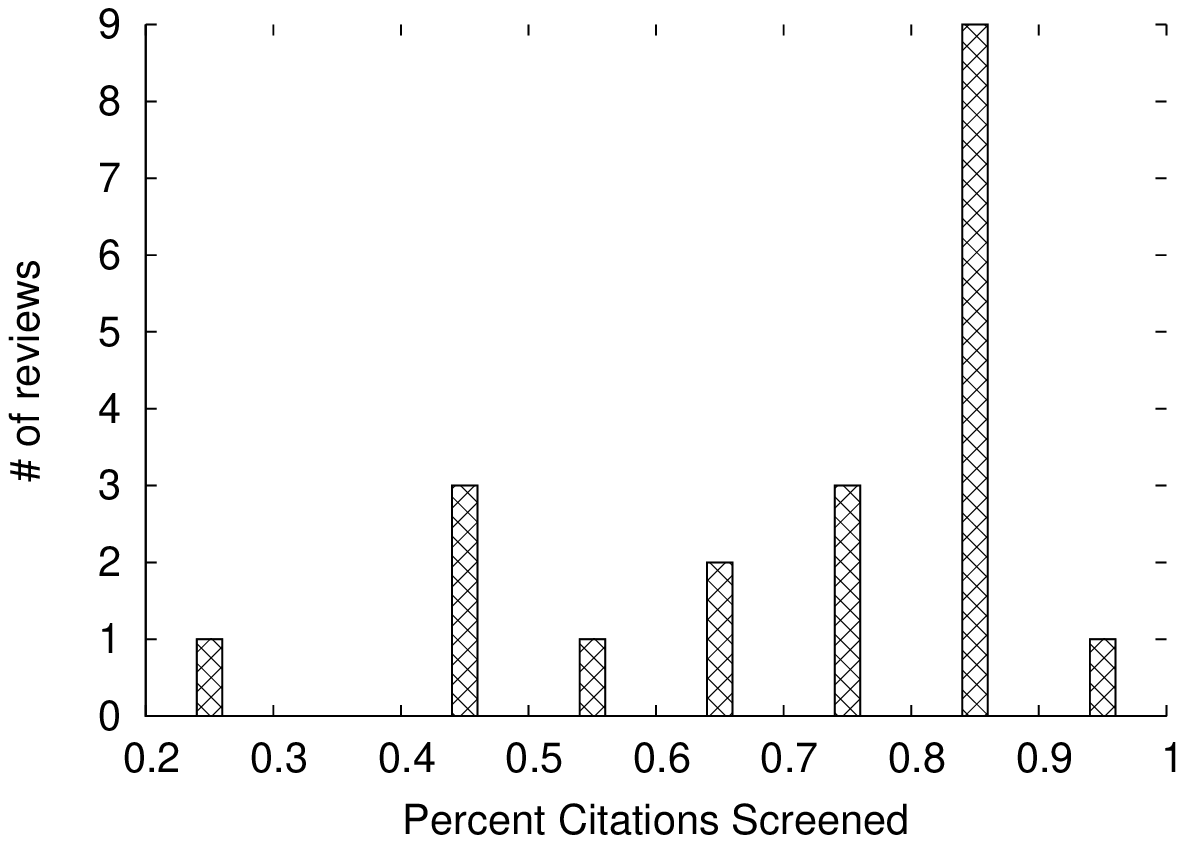}  &  
   \includegraphics[width=7cm]{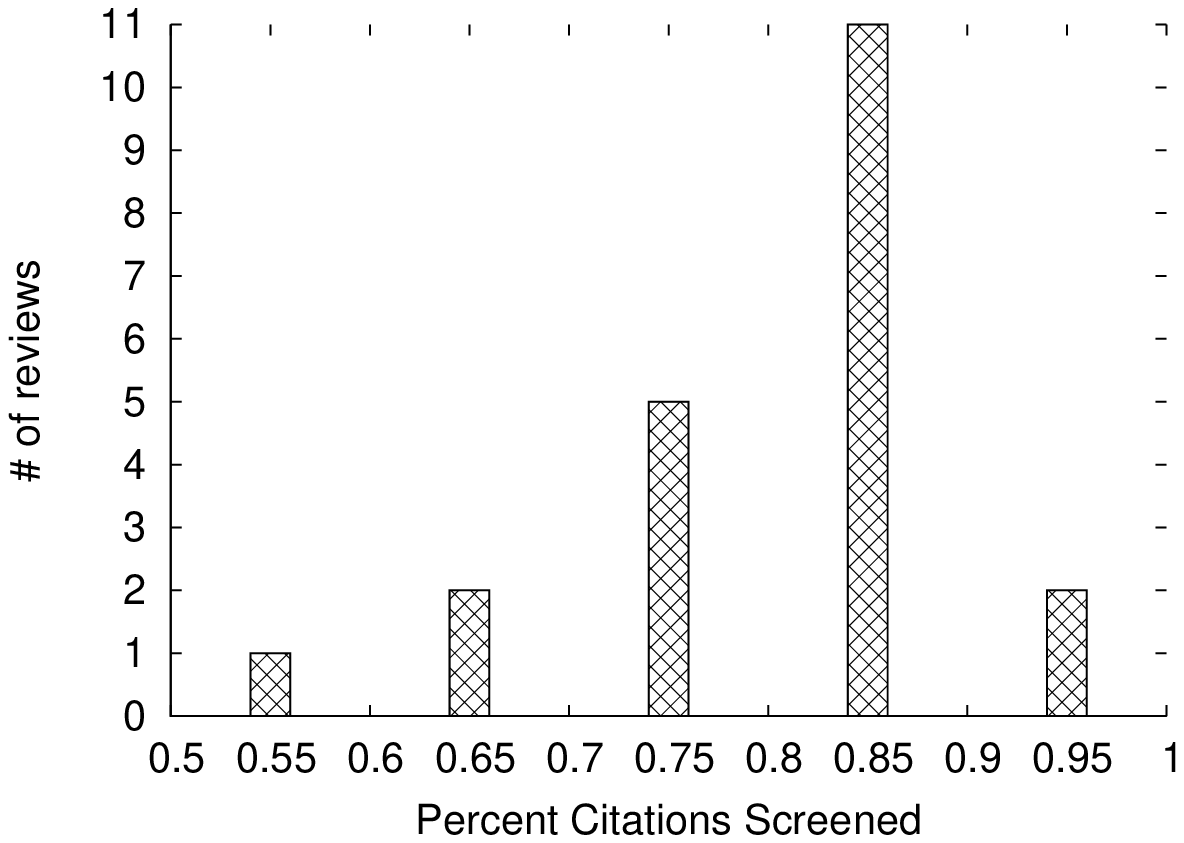} \\
\bottomrule
\end{tabular}
}
\caption{Active Learning experiment with a variable test set. The results are shown based on three groups of prevalence values. Prevalence is the ratio of the number of relevant citations out of the total citations. The X-axis represents the percentage of the total citations to screen for getting all relevant citations for a particular review. The Y-axis shows the total number of reviews.}
\label{tab:active_learning}

\end{figure}

\subsection{Experiment in an Active Learning Setting}
In this experiment, we use the $SVM^{perf}$ (b=1, AUC) ($21$) method as it is the top performing method in terms of the AUC metric in all three prevalence groups (Table~\ref{tab:rank-group}). 
For training, we choose $5$ relevant and $45$ irrelevant citations uniformly at random from the entire set and then learn a hyperplane $h$. We calculate the distance (score) from $h$ for each of the unlabeled instances and rank them based on this score. We choose the top-$50$ from the ranked citations to retrain the model along with the existing labeled citations. 
We repeat this experiment $500$ times and take the average. The goal is to see the following: if a particular user labels $50$ top ranked citations per batch, what is the percentage of total citations that the user has to screen to get all the relevant ones? Figure~\ref{tab:active_learning} shows the results of the proposed experiment. As before, the results are shown based on three groups of prevalence values. As shown in Figure \ref{tab:active_learning}, for the Low prevalence group, out of $20$ reviews, $7$ reviews (the first four values on the top histogram) need $40\%$ of the total citations to be screened to get all the relevant citations, $12$ reviews need around $50$ to $70\%$ citations, and $1$ review needs more than $90\%$ citations. 
For the mid and high prevalence groups, $9$ out of $20$ and $11$ out of $21$ reviews need around $80\%$ to $90\%$ citations to be screened to get all the relevant citations.

\begin{figure}[t]\centering
\centering
    \begin{subfigure}{0.45\textwidth}
        \centering
        \includegraphics[scale=0.40]{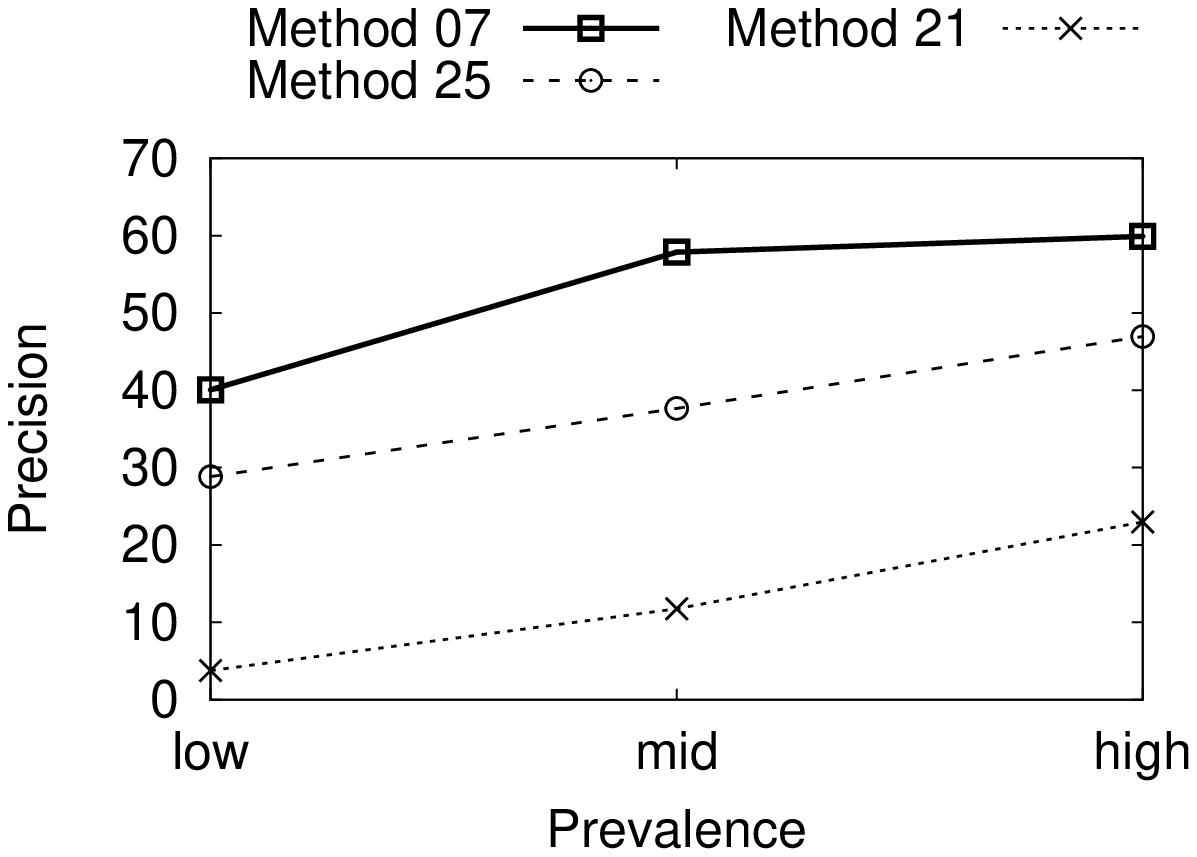} 
        \caption{} \label{fig:a}
    \end{subfigure}
    \begin{subfigure}{0.45\textwidth}
        \centering
        \includegraphics[scale=0.40]{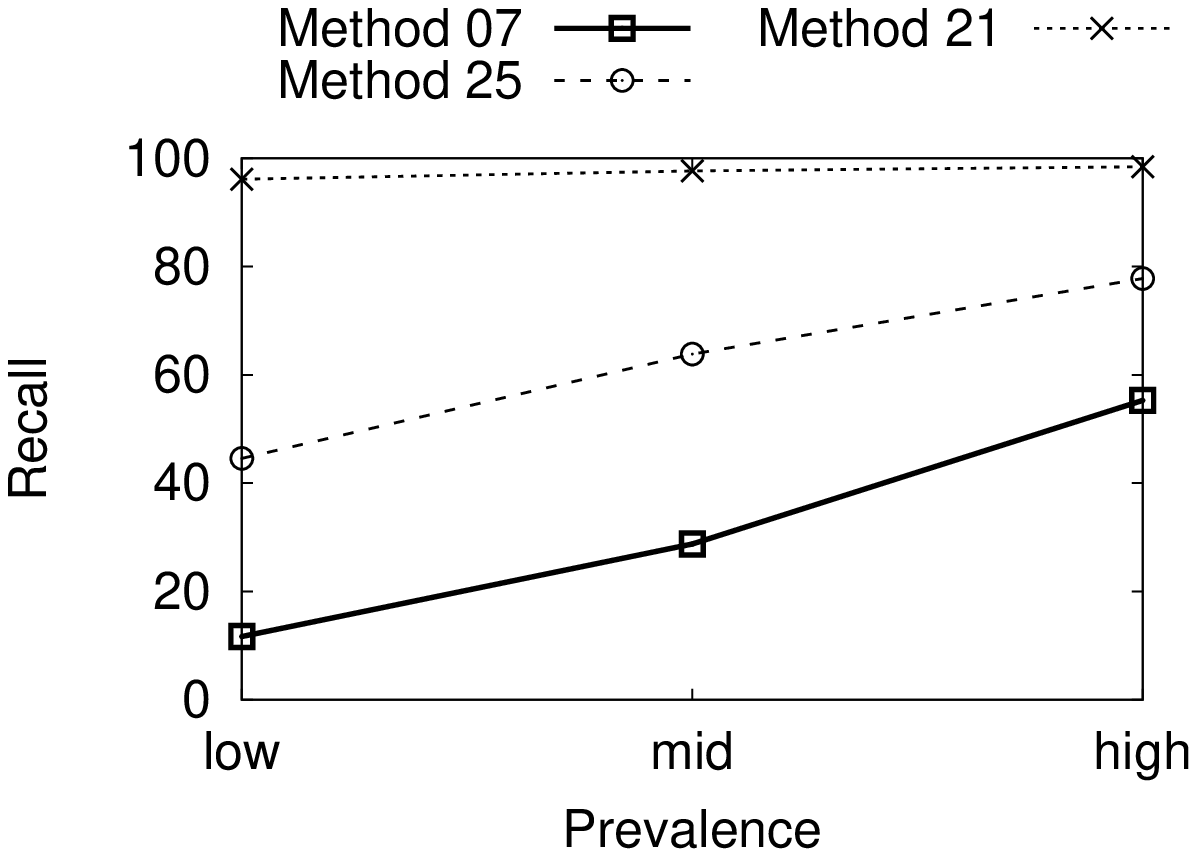} 
        \caption{} \label{fig:b}
    \end{subfigure}
   \begin{subfigure}{0.45\textwidth}
       \centering
       \includegraphics[scale=0.40]{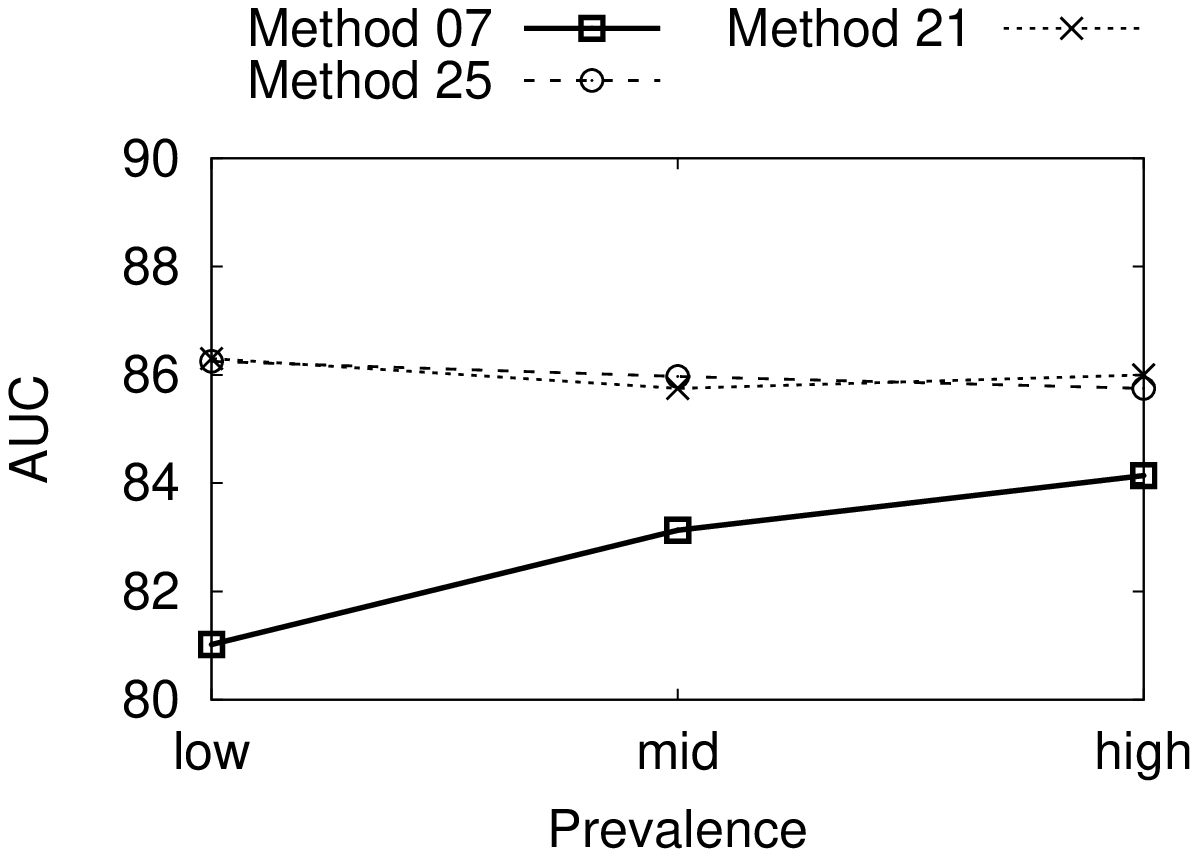} 
       \caption{} \label{fig:c}
   \end{subfigure}
   \begin{subfigure}{0.45\textwidth}
       \centering
       \includegraphics[scale=0.40]{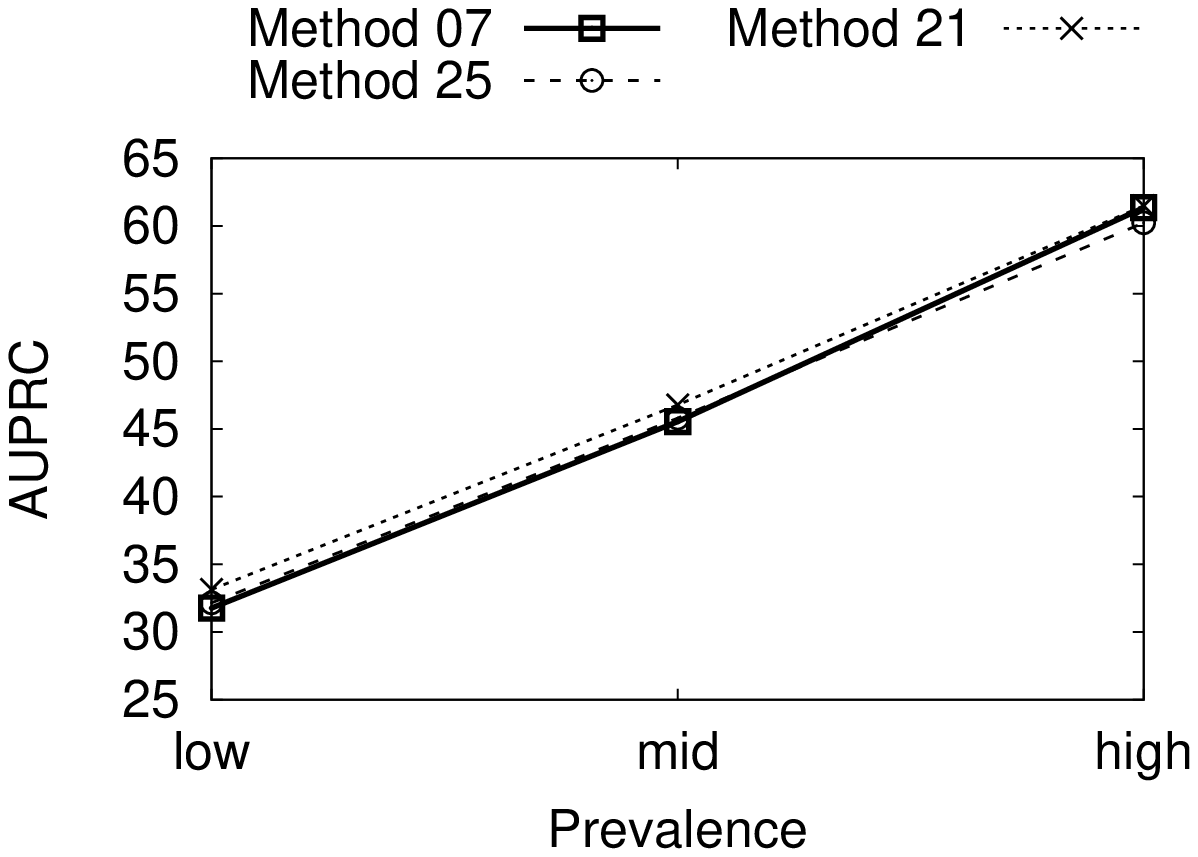} 
       \caption{} \label{fig:d}
   \end{subfigure}
\caption{\red{Performance of $SVM^{cost}$ (J, b=0) [Method 7], $SVM^{perf}$ (AUC, b=1) [Method 21], and $SVM^{cost}$ (J, b=1) [Method 25] in Precision, Recall, AUC, and AUPRC metric over three prevalence groups.}}
\label{fig:choice-of-algorithm}
\end{figure}

\subsection{Choice of Algorithms for \alg}
\label{sec:choice}
\red{Results in Section \ref{sec:result:existing} show that there is not a single algorithm which wins over all the metrics and prevalence groups. 
Our aim here is to choose an ensemble of algorithms in a way that  captures as much Precision as possible while maintaining a reasonably good Recall as we go up in the starring threshold. 
For this, we choose $SVM^{cost}$ (J, b=0) [Method 7], $SVM^{perf}$ (AUC, b=1) [Method 21], and $SVM^{cost}$ (J, b=1) [Method 25]. 
The choice of these algorithms is guided by the characteristics that these algorithms  have in different metrics (please see next paragraph) 
and based on the generalized performance over different prevalence groups measured through a well defined statistical test.  
From Table \ref{sec:result:existing}, we observe that Method 7 has the best performance for Precision in low and mid prevalence group and also performs reasonably good in Burden metric. 
Both Methods 5 and 7 are the candidates here; however, we choose Method 7 because the algorithm takes into account the data imbalance ratio (through the J parameter of the algorithm). Method 21 performs the best in many different metrics such as Recall, AUC, AUPRC, Yield, and Utility. 
Method 25 also performs the best across many different metrics (see Table \ref{sec:result:existing}) such as F-Measure, AUC, AUPRC, AM ER., and QD ER.}

\red{Figure \ref{fig:choice-of-algorithm} shows the generalized performance of the three algorithms used in our  ensemble method over three different prevalence groups in 
terms of Precision, Recall, AUC, and AUPRC. 
In AUPRC, all the three methods showed an increasing performance as the prevalence increases. 
For Method 21 and 25, AUC is consistent over different prevalence groups. 
Method 7 has a rising trend in AUC as the prevalence increases. 
For Recall, Method 21 has surpassed the other two methods by a wide margin.
Thus, we consider this algorithm as the dominant classifier. 
The other two methods showed an increasing trend as the prevalence of the dataset increases. 
On the other hand, for Precision, Method 7 has the highest Precision, and the Precision value increases with prevalence of the dataset  (see Figure \ref{fig:choice-of-algorithm}(a)) increases. Two other methods have much lower Precision than Method 7; however, their Precision is also increasing as the dataset prevalence increases.}


\begin{table}[!t]
\centering
\resizebox{0.75\textwidth}{!}{
\begin{tabular}{@{}lp{2.0cm}p{2.2cm}p{2.2cm}@{}}\toprule
{\bf{Metric}} & {\bf{RelRank ($3$-star)}} & {\bf{RelRank ($4$-star)}} & {\bf{RelRank ($5$-star)}}\\\midrule 
Recall  &$\langle {\bf{1}},1,1 \rangle$ &$\langle {\bf{2}},2,2 \rangle$ & $\langle {\bf{3}},4,5 \rangle$\\
Precision  &$\langle 9,10,12 \rangle$  &$\langle 8,9,10 \rangle$  & $\langle {\bf{7}},7,7 \rangle$ \\
F-Measure &$\langle 7,7,9\rangle$  &$\langle 6,6,7 \rangle$  & $\langle 5,4,4\rangle$ \\
Accuracy  & $\langle 9,13,11\rangle$ & $\langle 8,11,9\rangle$ & $\langle 7,9,6\rangle$ \\
\midrule
ROC (AUC) &$\langle 1,1,1 \rangle$  &$\langle 1,1,1 \rangle$  &$\langle 1,1,1 \rangle$\\
AUPRC  & $\langle 1,1,1\rangle$ &$\langle 1,1,1 \rangle$  & $\langle 1,1,1 \rangle$\\
\midrule
AM ER. &$\langle 5,7,8\rangle$  & $\langle 3,5,4\rangle$ & $\langle 1,1,1\rangle$\\ 
QD ER. &$\langle 5,9,9\rangle$  &$\langle 3,6,6\rangle$  &$\langle 1,1,1\rangle$ \\
\midrule
Burden &$\langle 11,14,13\rangle$  &$\langle 10,13,11\rangle$  & $\langle 8,10,8\rangle$\\
Utility &$\langle 1,1,1\rangle$  &$\langle 2,2,2\rangle$ &$\langle 3,4,5\rangle$ \\
\bottomrule
\end{tabular}}
\caption{Performance of \alg\ in different starred thresholds. The lower is the rank, the better is the performance. The three numbers inside the brackets indicate rank group for $\langle$Low, Mid, High$\rangle$ prevalence group. 
}
\label{tab:five-star-result}
\end{table}

\begin{table}[!t]
\centering
\resizebox{0.97\textwidth}{!}{
\begin{tabular}{@{}lccc@{}}\toprule
\textbf{Metric} & {\bf{RelRank ($3$-star)}} & {\bf{RelRank ($4$-star)}} & {\bf{RelRank ($5$-star)}} \\
\midrule
Precision & $\langle$ 03.78, 11.78, 22.98 $\rangle$ & $\langle$ 04.63, 15.56, 31.43 $\rangle$  & $\langle$ 06.22, 21.77, 41.62 $\rangle$ \\ 
Recall & $\langle$ 96.11, 97.65, 98.43 $\rangle$ & $\langle$ 92.40, 94.16, 93.82 $\rangle$  & $\langle$ 85.37, 86.50, 84.10 $\rangle$ \\ 
AUC & $\langle$ 86.38, 86.70, 86.34 $\rangle$ & $\langle$ 86.70, 86.70, 86.34 $\rangle$ &  $\langle$ 86.38, 86.70, 86.34 $\rangle$ \\
AUCPR & $\langle$ 36.27, 49.87, 63.81 $\rangle$ & $\langle$ 36.27, 49.87, 63.81 $\rangle$ & $\langle$ 36.27, 49.87, 63.81  $\rangle$ \\
Burden & $\langle$ 90.06, 91.10, 93.03 $\rangle$ & $\langle$ 78.29, 78.80, 78.89 $\rangle$ & $\langle$ 68.24, 68.66, 69.15 $\rangle$ \\
Utility & $\langle$ 93.64, 94.32, 94.60 $\rangle$ & $\langle$ 92.38, 93.28, 93.12 $\rangle$ & $\langle$ 89.63, 90.15, 88.99 $\rangle$ \\
Yield & $\langle$ 98.04, 98.82, 99.21 $\rangle$ & $\langle$  96.10, 97.08, 96.91 $\rangle$ & $\langle$ 92.67, 93.25, 92.05 $\rangle$ \\
AM Error & $\langle$ 0.417, 0.414, 0.424 $\rangle$ & $\langle$ 0.316, 0.298, 0.276  $\rangle$ & $\langle$ 0.248, 0.228, 0.216 $\rangle$ \\
QM Error & $\langle$ 0.517, 0.573, 0.591 $\rangle$ & $\langle$ 0.407, 0.385, 0.353  $\rangle$ & $\langle$ 0.286, 0.283, 0.230 $\rangle$ \\
\bottomrule
\end{tabular}}
\caption{\red{Performance values of \alg\ in different starred thresholds. 
For Precision, Recall, AUC, AUCPR, Burden, Utility, and Yield, the higher the value, the better  the performance. On the other hand, for AM and QM errors, the lower the better. The three numbers inside the brackets indicate the value of the metric for $\langle$low, mid, and high$\rangle$ prevalence groups.}}
\label{tab:relrank-3-perf}
\end{table}

\subsection{Performance of our Proposed 5-star Rating Algorithm}
For a SR platform, we need easy to consume prediction, $i.e.$, the citations should be sorted based on their relevance (discussed in Introduction). The most important metrics in our case are: AUC, AUPRC, Recall and Utility. We evaluate our $5$-star algorithm in different cumulative star-groups. Table~\ref{tab:five-star-result} shows the obtained results. \red{We also present the average performance of a metric in different prevalence groups and starring thresholds in Table \ref{tab:relrank-3-perf}.} We observe that our $5$-star algorithm performs similarly to method $21$ for the citations receiving $3$ stars or above. For the Precision metric, the 5-star algorithm performs better than method 21 as it falls in the rank group 7 whereas method 21 falls in the group 9. This is because  our algorithm is a combination of method $21$ with two other methods that have better Precision performances. 
We see an increment of around 10\% in Precision in the high prevalence group from \alg\ (3-star) to \alg\ (4-star) and then to \alg\ (5-star); the corresponding values for Precision are 22.98\%, 31.43\%, and 41.62\%, respectively. For the mid prevalence group, the increment is around 4-6\% and for the low one, it is around 2\%.
Similarly, for the Recall metric, \alg\ (3-star) is in the top group whereas \alg\ (4-star) and \alg\ (5-star) are in the second top group. As the Precision goes high, Recall decreases as we go up in the rating. For the high prevalence group, it decrements from 98.43\% to 93.82\% and then to 84.10\%. For the low and mid prevalence groups, we also see a similar decrement of 3-7\% in performance.

Interestingly, for both AUC and AUPRC, \alg\ falls in the top group (see Table \ref{tab:rank-group} and \ref{tab:relrank-3-perf} to observe that this is not the case if the algorithms are used alone) which makes our ensemble algorithm more suitable for SR applications. Our algorithm achieves around 87\% AUC in all prevalence groups and around 37\%, 50\% and 64\% AUPRC for the low, mid and high prevalence groups, respectively. For Utility, we also see a similar behavior. 
Burden drops from around 90\% to 78\% and then to 68\%, going from \alg\ (3-star) to \alg\ (4-star) and then to \alg\ (5-star) in all prevalence groups as expected. 
However, Yield which represents the fraction of citations that are identified by a given screening approach (i.e., \alg\ ) does not drop much as we go up in the star rating. It only drops around 2-4\% which is a very important aspect of our algorithm as the rate \alg\ is finding relevant citation is not decreasing so rapidly.
For AM and QD errors, which take into account loss in Recall in both the relevant and irrelevant classes, we see a drop around 10-12\% from \alg\ (3-star) to \alg\ (4-star) to \alg\ (5-star). This is because of the Precision/Recall trade-off. In \alg\ (3-star), we achieve the highest Recall and the lowest Precision in the positive class which indicates that we loose Recall in the irrelevant class. This is why the AM and QM errors are very high (AM error is around 40\%, QM error around 50\%) in the \alg\ (3-star).

\section{Discussion} \label{sec:discussion}

In our study of different methods, 
we observed that almost always there is a method that ranks first in the three prevalence groups. 
However, there is no single dominant method across all metrics. 
Various methods perform well on different prevalence groups and for different metrics. 
For instance, w2v row with $SVM^{perf}$ (b=1, AUC) ($21$) seems to be a good choice, outperforming the other methods in five metrics. The method achieves around 97\%  Recall and 87\% AUC across different prevalence groups and datasets. 
However, it is not present in any of the equivalence groups for a few other metrics. The method gives around 4\% Precision in low prevalence group, 11.75\% in mid-prevalence group and 23\% in high prevalence group. The same behavior is seen for other metrics in which it performed poorly. This is because this method has a high Recall, but a low Precision. As a consequence, such a method is penalized heavily by some metrics such as AUPRC and Burden. This comprehensive study and subsequent analysis on our abstract screening platform Rayyan suggests that a holistic ``composite'' strategy is a better choice in a real-life abstract screening system.

We proposed such a composite method, called \alg\ in Section~\ref{sec:5-star} and showed that it performs well across many metrics. This is not surprising since each of the baseline methods has been designed to meet a certain objective. However, abstract screening is a complicated process in which a ``good" method should be able to optimize simultaneously several metrics such as Recall, AUC, and Utility, which can be achieved by a composite strategy like \alg. 
We also see a similar behavior while doing a manual verification over a set of seven reviews. 
As  Table~\ref{tab:five-star-result} shows, we fall short on some other metrics, these are the metrics that directly depend on Precision. However, from our user surveys, we realized that if the ranking is very good then a user can generally exclude many citations from the bottom without so much effort and the prediction actually makes more sense. So, we strictly emphasize on ranking in our combined algorithm.

\begin{wrapfigure}{r}{4.5cm}
\centering
\includegraphics[scale=0.40]{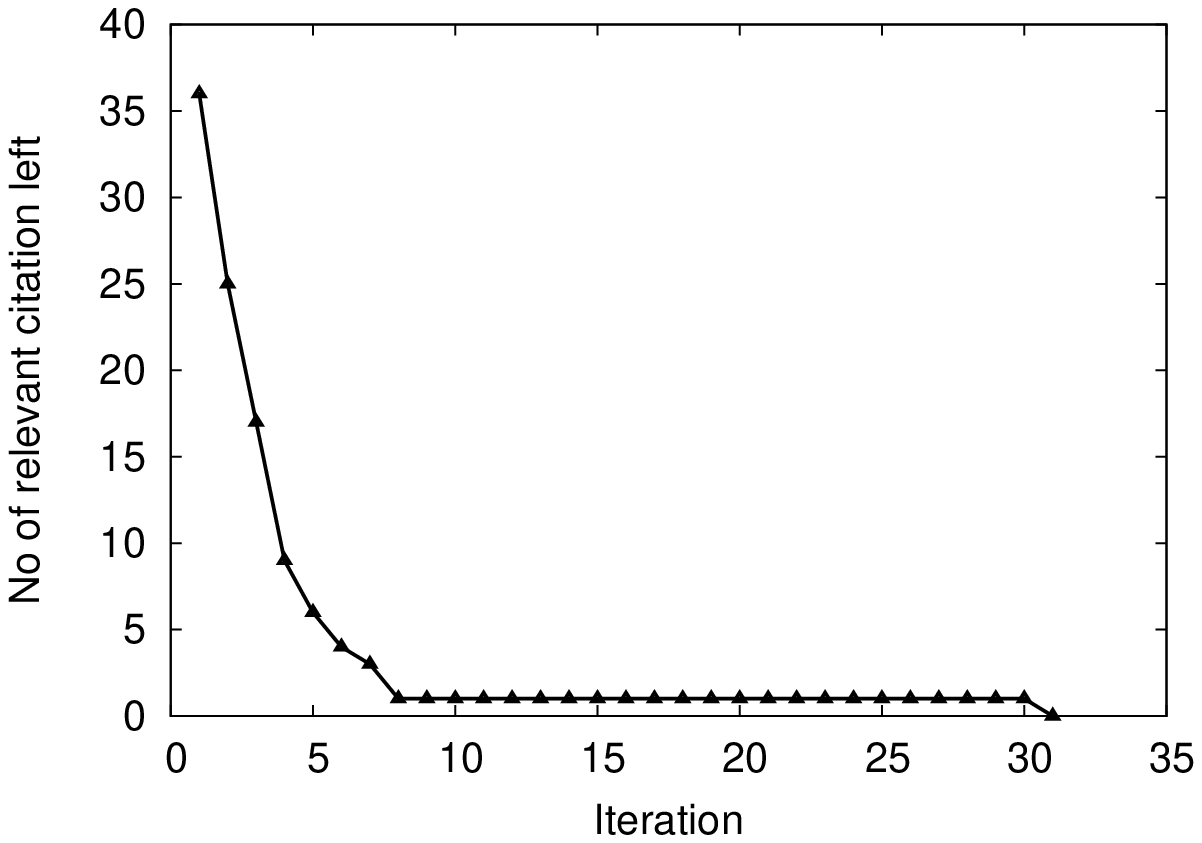}
\caption{The pattern of decrease in the number of relevant citations during abstract screening 
for a particular review.}
\label{fig:inclusion}
\end{wrapfigure}

We also analyzed the performance of an abstract screening system in an active learning setting. 
We observe that the 3-prevalence groups exhibit similar characteristics, 
$i.e.$, some of the citations add the burden of reviewing almost all of them. In a separate experiment (in addition to the reported experiments in the results section), we further analyze the inclusion behavior of a particular random run of review $1$. \red{Figure~\ref{fig:inclusion} shows the results of this experiment.} For a random run, around $58\%$ ($1500$ out of $2544$) of the documents have to be screened before it gets its final relevant citation. However, by getting into the details of the experiment, we see that it gets almost all but the final one after screening only $400$ out of $2544$ which is around $15\%$ of the total citations. The last citation surprisingly takes $22$ more iterations. With the help of a domain expert, it would be interesting to see whether the final citation has outlying characteristics. We see a similar behavior for some other reviews. In our opinion, in addition to capturing the relevancy information, a holistic model for abstract screening may also have another component to exclude and point out outlying citations which we keep as our future work.

\subsection{Limitations of our Study} \label{sec:discuss:limit}

Our study is specifically targeted to study and improve abstract screening in a practical systematic review platform like Rayyan which serves thousands of users. Therefore, this study is quite different from the standalone evaluation of single abstract screening tasks as the method for abstract screening has to generalize over thousands of reviews from thousands of users. For this reason, we have carefully avoided some of the feature representation techniques such as  co-citations and MeSH (Medical Subject Heading) which are not readily available even though some of the previous works have shown better performances with those features. 
Furthermore, according to the study \cite{Olorisade.De:16}, many methods other than Support Vector Machines (SVM) are used as classification algorithms for abstract screening. However, in this study, we restricted our attention to only SVM-based methods. Our proposed 5-star method also combines three of the SVM methods. One of the reasons for this choice from the system perspective is that SVM methods require a small amount of memory to store models and hence allows for storage savings. For example, for $SVM^{perf}$, we need to store only the feature importance values.


\section{Conclusion} \label{sec:conclusion}
In this paper, we studied the most popular classification methods employed in abstract screening for systematic review. We focused on the algorithms that better fit the constraints of a  real-world system like Rayyan. We found that there is no single ``winner" approach, $i.e.$, various methods performed well on different prevalence groups and for different metrics. For instance, w2v row with $SVM^{perf}$ (b=1, AUC) outperformed all the other studied methods in five metrics but not on few others.
We also observe that in an active learning setting, a substantial portion of included citations is discovered within a few iterations. However, one or two citations had some outlying behavior, and the method would require more iterations to retrieve these citations. We also presented an ensemble method, namely \alg\, that combines three of the studied methods. Our approach converts their scores into a 5-star rating system through a voting mechanism and ranks citations efficiently based on their graded relevance.

\bibliographystyle{elsarticle-harv}
\bibliography{sysrev}

\end{document}